\documentclass[12pt,a4paper]{iopart}
\pdfoutput=1
\usepackage{iopams,setstack,amsthm,cite,float,graphicx,slashed,bbold}
\usepackage[shortlabels]{enumitem}
\setlist[enumerate,1]{itemsep=2pt,parsep=0pt,topsep=6pt,leftmargin=2\parindent}
\usepackage[utf8]{inputenc}
\usepackage[T1]{fontenc}
\usepackage{lmodern}
\usepackage[colorlinks,linkcolor=blue,citecolor=blue,urlcolor=blue]{hyperref}
\newcommand{\al}{\alpha}
\newcommand{\be}{\beta}
\newcommand{\de}{\delta}

\newcommand{\vep}{\varepsilon}
\newcommand{\ga}{\gamma}

\newcommand{\la}{\lambda}

\newcommand{\si}{\sigma}
\renewcommand{\th}{\theta}
\newcommand{\vp}{\varphi}

\newcommand{\ze}{\zeta}
\newcommand{\De}{\Delta}

\newcommand{\La}{\Lambda}

\newcommand{\bn}{\mathbf{n}}

\newcommand{\bbs}{\mathbf{s}}

\newcommand{\bx}{\mathbf{x}}

\newcommand{\bsi}{{\boldsymbol{\si}}}
\newcommand{\bz}{\mathbf{z}}

\newcommand{\bDe}{\boldsymbol{\De}}
\newcommand{\tS}{\widetilde{S}}

\newcommand{\tn}{\widetilde{n}}

\newcommand{\tnu}{\widetilde\nu}

\newcommand{\CC}{{\mathbb C}}
\newcommand{\NN}{{\mathbb N}}
\newcommand{\RR}{{\mathbb R}}
\newcommand{\ZZ}{{\mathbb Z}}
\newcommand{\cE}{{\mathcal E}}

\newcommand{\cH}{{\mathcal H}}

\newcommand{\cP}{{\mathcal P}}

\newcommand{\cS}{{\mathcal S}}
\newcommand{\cT}{{\mathcal T}}

\newcommand{\cZ}{{\mathcal Z}}
\newcommand\Hsc{H_{\mathrm{sc}}}
\newcommand\Hsp{H_{\mathrm{spin}}}
\newcommand\Zsc{Z_{\mathrm{sc}}}

\newcommand{\pd}{\partial}

\newcommand{\id}{\mathbb 1}

\newcommand{\ket}[1]{|#1\rangle}
\newcommand{\bra}[1]{\langle#1|}

\let\ds\displaystyle

\newcommand{\mss}{\kern 1pt}

\renewcommand{\le}{\leqslant}
\renewcommand{\ge}{\geqslant}
\newcommand{\tends}[1]{\bbuildrel{\hbox to 2em{\rightarrowfill}}_{#1}^{}}

\newcommand{\operatorname}[1]{\mathop{\rm #1}\nolimits}
\newcommand{\sech}{\operatorname{sech}}
\newcommand{\csch}{\operatorname{csch}}

\newcommand{\str}{\operatorname{str}}

\newcommand{\iu}{\mathrm i}
\newcommand{\diff}{\mathrm{d}}

\newcommand{\su}{\mathrm{su}}

\newcommand{\sla}{\mathrm{sl}}
\newcommand{\gl}{\mathrm{gl}}

\newcommand{\implies}{\Longrightarrow}

\newcommand{\en}{\enspace}

\newcommand{\Int}[1]{\,\mathop{\!#1}\limits^{\lower1ex\hbox{$\scriptstyle\circ$}}{}}

\newcommand{\HS}{\mathrm{HS}}

\newcommand{\dc}{c^\dagger}

\newcommand{\abs}[1]{\vert #1\vert}
\newcommand{\Abs}[1]{\bigl\vert #1\bigr\vert}

\theoremstyle{remark}
\newtheorem{remark}{Remark}

\let\tfrac\case
\let\eqref\eref

\newcommand{\binom}[2]{{#1\choose #2}}
\newcommand\vac\emptyset
\eqnobysec
\newcommand{\dGS}{d_\mathrm{GS}}
\newcommand{\EGS}{E_\mathrm{GS}}
\def\clap#1{\hbox to 0pt{\hss#1\hss}}

\newcommand{\sumnhalf}{\su\bigl(\frac{m}2|\frac{n}2\bigr)}
\begin{document}

\title[Novel translationally invariant supersymmetric chain]{A novel translationally invariant
  supersymmetric\\ chain with inverse-square interactions: partition function, thermodynamics and
  criticality}

\author{Bireswar Basu-Mallick, Federico Finkel, Artemio González-López }

\address{Depto.~de Física Teórica, Facultad de Ciencias Físicas, Plaza de las Ciencias 1,\\
  Universidad Complutense de Madrid, 28040 Madrid, SPAIN}

\eads{\mailto{bireswar.basumallick@saha.ac.in}, \mailto{ffinkel@ucm.es}, \mailto{artemio@ucm.es}}

\vspace{10pt}
\begin{indented}
\item[]{}September 16, 2024
\end{indented}
\begin{abstract}
  We introduce a novel family of translationally-invariant $\su(m|n)$ supersymmetric spin chains
  with long-range interaction not directly associated to a root system. We study the symmetries of
  this model, establishing in particular the existence of a boson-fermion duality characteristic
  of this type of systems. Taking advantage of the relation of the new chains with an associated
  many-body supersymmetric spin dynamical model, we are able to compute their partition function
  in closed form for all values of $m$ and $n$ and for an arbitrary number of spins. When both $m$
  and $n$ are even, we show that the partition function factorizes as the product of the partition
  functions of two supersymmetric Haldane--Shastry spin chains, which in turn leads to a simple
  expression for the thermodynamic free energy per spin in terms of the Perron eigenvalue of a
  suitable transfer matrix. We use this expression to study the thermodynamics of a large class of
  these chains, showing in particular that the specific heat presents a single Schottky peak at
  approximately the same temperature as a suitable $k$-level model. We also analyze the critical
  behavior of the new chains, and in particular the ground state degeneracy and the existence of
  low energy excitations with a linear energy-momentum dispersion relation. In this way we show
  that the only possible critical chains are the ones with $m=0,1,2$. In addition, using the
  explicit formula for the partition function we are able to establish the criticality of the
  $\su(0|n)$ and $\su(2|n)$ chains with even $n$, and to evaluate the central charge of their
  associated conformal field theory.
\end{abstract}

\noindent {\it Keywords\/}: integrable spin chains and vertex models; solvable lattice models;
quantum criticality.


\maketitle


\section{Introduction}
\label{sec.intro}

In our previous paper~\cite{BFG20} we introduced a novel translationally invariant spin chain with
inverse-square interactions depending on both spin permutation and spin reversal operators, which
reduces to the celebrated Haldane--Shastry (HS) chain~\cite{Ha88,Sh88} when the latter operators
are replaced by plus or minus the identity. The new model, which consists of $N$ sites each of
which is occupied by particles of a single species transforming under the fundamental
representation of the $\su(m)$ Lie algebra, is exactly solvable, in the sense that its partition
function can be evaluated in closed form for arbitrary $N$. The ultimate reason for this is that
the spin chain can be obtained as the strong coupling limit of a one-dimensional many-body spin
dynamical model, whose partition function can be computed in closed form in this limit. The
chain's partition function can then be evaluated applying Polychronakos's freezing
trick~\cite{Po93,Po94}, which basically amounts to modding out the dynamical degrees of freedom of
the spin dynamical model.

A remarkable property of the new solvable chain introduced in Ref.~\cite{BFG20} is the fact that,
unlike what is the case with practically all integrable spin chains with long-range interactions,
it is not directly associated to a single classical (extended) root system (see, e.g.,
\cite{OP83,CS02,OS04}). More precisely, since the latter chain is translationally invariant the
interaction between two sites depends only on their distance, as is the case for spin chains of HS
type constructed from the $A_{N-1}$ root system. On the other hand, the chain's Hamiltonian
includes not only spin permutation but also spin reversal operators, in a combination
characteristic of spin chains related to the $D_N$ root system.

The structure of the new chain's partition function turns out to be particularly simple in the
case of even $m$. Indeed, in this case the model's Hamiltonian is unitarily equivalent to the sum
of two independent HS chain Hamiltonians, an ordinary one of type $\su\bigl(\frac{m}2\bigr)$ and a
supersymmetric one of $\su(1|1)$ type. This yields an elegant description of the spectrum in terms
of Haldane's (supersymmetric) motifs~\cite{HHTBP92,Ha93,HB00,BBHS07} of $\su\bigl(\frac{m}2\bigr)$
and $\su(1|1)$ types, and also establishes the invariance of the model under the direct sum of the
\hbox{(super-)Yangians} $Y(\gl(0|m/2))$ and $Y(\gl(1|1))$. In fact, the general $\su(m|n)$
supersymmetric version of the original HS chain was introduced early on by Haldane
himself~\cite{Ha93}. This model, which consists of two species of particles behaving as bosons and
fermions with $m$ and $n$ internal degrees of freedom, respectively, was thoroughly studied in
Ref.~\cite{BB06}. In particular, a closed-form expression for its partition function was derived
in the latter reference, which was used in turn in Ref.~\cite{BBH10} to obtain a complete
description of the spectrum in terms of $\su(m|n)$-supersymmetric bond vectors and their
associated motifs.

In view of the above, it is natural to consider the $\su(m|n)$-supersymmetric generalization of
the translationally invariant spin chain of Ref.~\cite{BFG20}. This is, indeed, the first aim of
this paper. More precisely, we shall construct the latter chain from its associated spin dynamical
model and evaluate its partition function in closed form for all values of $m$ and $n$. We shall
show that when both $m$ and $n$ are even the model's Hamiltonian is unitarily equivalent to the
sum of the Hamiltonians of two supersymmetric HS chains of $\su(1|1)$ and $\sumnhalf$ types. As in
the non-supersymmetric case, this automatically entails a simple description of the chain's
spectrum in terms of $\su(1|1)$ and $\sumnhalf$ supersymmetric motifs.

The thermodynamics of the original HS chain and its rational and hyperbolic variants has been
extensively studied in the literature since its very inception~\cite{Ha91,Fr93,FI94}. More
recently, the description of the spectrum in terms of Haldane motifs was systematically used in
Ref.~\cite{EFG12} to derive a closed-form expression of the thermodynamic functions of all
(non-supersymmetric) spin chains of HS type related to the $A_{N-1}$ root system. This method was
later extended to $\su(m|n)$ supersymmetric chains of HS type in Ref.~\cite{FGLR18}. Our second
aim is to take advantage of the motif-based description of the spectrum of the $\su(m|n)$
supersymmetric chain introduced in this paper when both $m$ and $n$ are even to compute the free
energy of this model in the thermodynamic limit, deriving closed-form expressions for its main
thermodynamic functions. In particular, using these expressions we shall show that the specific
heat per spin exhibits a single Schottky peak, whose temperature is close to the temperature of
the Schottky peak of an appropriate $k$-level system. When either $m$ or $n$ is odd, we shall use
the exact expression of the partition function derived in this work to study the thermodynamic
functions for $N$ finite but as large as possible. We shall show that, barring finite-size effects
at very low temperatures, their behavior is qualitatively analogous to that of their counterparts
when both $m$ and $n$ are even.

One-dimensional spin chains are one of the simplest systems exhibiting quantum phase
transitions~\cite{Sa11}. This typically occurs when the spectrum is gapless in the thermodynamic
limit, and the model's low-energy sector can be effectively described by a suitable
($1+1$)-dimensional CFT (or, if the ground state is degenerate, by several copies thereof).
%
%
For this to be possible the system's ground state must have finite degeneracy in the thermodynamic
limit, and it must possess low-energy excitations above the ground state exhibiting a linear
energy-momentum relation with a characteristic Fermi velocity $v_F$. This is the case, for
example, for the original (antiferromagnetic) $\su(m)$ HS chain~\cite{HHTBP92,SC94,BS96}, whose
low-energy excitations are known to be governed by the $\su(m)_1$ Wess--Zumino--Novikov--Witten
model~\cite{PW83,Wi84,KZ84}. The critical behavior of $\su(m|n)$-supersymmetric HS spin chains was
also analyzed in Ref.~\cite{BBS08}, where it was established that only the $\su(1|m)$ chain (i.e.,
with one bosonic and $n$ fermionic degrees of freedom) is also critical. Another hallmark of
critical quantum systems is the low temperature behavior of their free energy, which should behave
as the free energy of a $(1+1)$-dimensional CFT. In other words, at low temperatures we should
have~\cite{BCN86,Af86}
\begin{equation}\label{fcrit}
  f(T)=f(0)-\frac{\pi cT^2}{6v_F}+o(T^2),
\end{equation}
where $c$ is the central charge of the associated CFT. In particular, the analysis of the
low-temperature behavior of the free energy of a critical quantum system provides an efficient way
of computing the central charge of its associated CFT, which in turn determines its universality
class. Our third objective is to analyze the critical behavior of the $\su(m|n)$ supersymmetric
chains introduced in this paper, and in particular to determine for what values of $m$ and $n$
they are critical and what is their central charge. By studying the ground state degeneracy of
these models, we shall show that they can only be critical for $m=0,1,2$. We shall then use the
motif-based description of the spectrum to prove that the $\su(0|n)$ and $\su(2|n)$ chains are
critical for all even values of $n$. This will be confirmed by the analysis of the low-temperature
behavior of the thermodynamic free energy per spin, which also yields the central charge of the
associated CFTs.

The rest of the paper is organized as follows. In Section~\ref{sec.model} we introduce the novel
$\su(m|n)$ spin chains under consideration, and explain how they can be obtained from a suitable
(supersymmetric) spin dynamical model in the strong coupling limit. Exploiting this connection, in
Section~\ref{sec.PF} we evaluate in closed form the chain's partition function for arbitrary $m$,
$n$ and $N$, and use the explicit expression thus obtained to study the ground state degeneracy.
In Section~\ref{sec.symm} we analyze the the new chain's main symmetries, showing that they
possess a remarkable ``twisted'' translation invariance as well as a boson-fermion duality
characteristic of this type of models. Section~\ref{sec.thermo} is devoted to the study of the
chain's thermodynamics. Taking advantage of the simple structure of the partition function, we are
able to find analytic expressions for the thermodynamic free energy per spin and the main
thermodynamic functions when $m$ and $n$ are both even. We also study the behavior of these
functions, showing in particular that the specific heat features a single Schottky peak. In
Section~\ref{sec.CB} we examine the critical behavior of the supersymmetric chains under study,
deriving the partial results explained above. We present our conclusions and outline several paths
for future research in Section~\ref{sec.conc}. The paper ends with two technical appendixes, in
which we discuss the precise connection of the $\su(m|n)$ Lie superalgebra with our model and
derive an asymptotic approximation for an integral used to ascertain the low-temperature behavior
of the free energy per spin of the critical chains.

\section{The model}
\label{sec.model}

The model we shall deal with in this paper is the supersymmetric version of the spin chain
introduced in Ref.~\cite{BFG20}. It describes a one-dimensional array of $N$ spins, each of which
can be either a boson or a fermion, lying on the upper unit half-circle at uniformly spaced
positions $\ze_k=\e^{2\iu\th_k}$, with
\begin{equation}\label{roots}
  \th_k:=\frac{k\pi}{2N},\qquad 1\le k\le N.
\end{equation}
More precisely, we shall suppose that there are $m$ bosonic and $n$ fermionic degrees of freedom,
so that the Hilbert space of the system is $\cS^{(m|n)}=\otimes_{i=1}^N\cS_i^{(m|n)}$ with
$\cS_i^{(m|n)}=\CC^{m+n}$. The canonical basis in this space shall be denoted by
\begin{equation}\label{basis}
  |s_1\cdots s_N\rangle\equiv\ket{\bbs}:=|s_1\rangle\otimes\cdots\otimes|s_N\rangle,
  \qquad 1\le s_i\le m+n.
\end{equation}
We shall regard the basis states $\ket{s_i}$ with $s_i\in B:=\{1,\dots,m\}$ as bosonic, and those
with $s_i\in F:=\{m+1,\dots,m+n\}$ as fermionic. The model's Hamiltonian admits a simple
expression in terms of the $\su(m|n)$ supersymmetric spin permutation and spin flip operators,
whose definition we shall next recall.

The supersymmetric spin permutation operators $S_{ij}^{(m|n)}=S_{ji}^{(m|n)}$ (with $i<j$) are
defined by
\begin{equation}\label{spin-perm}
  S_{ij}^{(m|n)}\ket{\cdots s_i \cdots s_j\cdots}:=(-1)^{\nu(s_i,\dots,s_j)}
  \ket{\cdots s_j\cdots s_i\cdots}\,,
\end{equation}
where $\nu(s_i,\dots,s_j)$ is $0$ (respectively $1$) if $s_i,s_j\in B$ (respectively
$s_i,s_j\in F$), and is otherwise equal to the number of fermionic spins $s_k$ with
$i+1\le k\le j-1$. Note that $S_{ij}^{(m|0)}$ is a standard permutation operator~$P_{ij}$, while
$S_{ij}^{(0|n)}=-P_{ij}$. Likewise, the action of the spin flip operators
$S_i^{(m\vep_B|n\vep_F)}$ on the canonical basis vectors is given by
\begin{equation}
  \label{spin-rev}
  S_i^{(m\vep_B|n\vep_F)}\ket{\cdots s_i\cdots}:=\si(s_i)\ket{\cdots s_i'\cdots}\,,
\end{equation}
with
\begin{equation}\label{sigma}
  \si(s_i)= \cases{
    \vep_B,\quad &$s_i\in B$,\\
    \vep_F,\quad &$s_i\in F$, }
\end{equation}
where $\vep_B,\vep_F\in\{\pm1\}$ are two fixed signs and $s_i\mapsto s_i'$ is the ``spin flip''
(involution) defined by
\begin{equation}\label{prime}
  s_i'= \cases{
    m+1-s_i,\quad &$s_i\in B$,\\
    2m+n+1-s_i,\quad &$s_i\in F$. }
\end{equation}
The precise connection between the spin permutation and reversal operators just defined with the
Lie superalgebra $\su(m|n)$ is explained out in~\ref{app.sumn}. In terms of these operators, the
model's Hamiltonian is defined as
\begin{equation}
  \label{Hchain}
  \cH^{(m\vep_B|n\vep_F)}=\frac14\sum_{1\le i<j\le N}\left(\frac{1-S_{ij}^{(m|n)}}{\sin^2(\th_i-\th_j)}
    +\frac{1-\tS_{ij}^{(m\vep_B|n\vep_F)}}{\cos^2(\th_i-\th_j)}\right)\,,
\end{equation}
where we have set
\[
  \tS_{ij}^{(m\vep_B|n\vep_F)}:=S_{ij}^{(m|n)}S_{i}^{(m\vep_B|n\vep_F)}S_{j}^{(m\vep_B|n\vep_F)}.
\]
In what follows we shall also usually suppress the superindices $m$, $n$, $\vep_B$, and $\vep_F$
from all operators, writing simply $S_{ij}$, $S_i$, $\tS_{ij}$ and $\cH$.
\begin{remark}\label{rem.vep}
  Since the Hamiltonian~\eqref{Hchain} only contains products of two spin flip operators, it is
  clear that $\cH$ is invariant under the replacement $\vep_B\to-\vep_B$, $\vep_F\to-\vep_F$. For
  this reason, from now on we shall suppose without loss of generality that
  \[
    \vep_B=1,\qquad \vep_F=\pm1.
  \]
  \hspace*{\fill}\qed
\end{remark}
\begin{remark}
  When $m=0$ and $\vep_F=1$ the Hamiltonian~\eqref{Hchain} reduces to the Hamiltonian of the
  non-supersymmetric spin chain introduced in Ref.~\cite{BFG20}. On the other hand, if the spin
  flip operators $S_i^{(m|n)}$ are replaced by $\pm\id$ then $\tS_{ij}^{(m|n)}=S^{(m|n)}_{ij}$
  and~\eqref{Hchain} becomes
  \begin{eqnarray*}
    \frac14\sum_{1\le i<j\le N}\Big(\sin^{-2}(\th_i-\th_j)&
      +\cos^{-2}(\th_i-\th_j)\Big)\left(1-S_{ij}^{(m|n)}\right)\\
    &=\sum_{1\le i<j\le N}\sin^{-2}\left(\tfrac{(i-j)\pi}N\right)
      \left(1-S_{ij}^{(m|n)}\right)\equiv2\cH^{(m|n)}_{\text{HS}},
  \end{eqnarray*}
  where $\cH^{(m|n)}_{\text{HS}}$ denotes the Hamiltonian of the $\su(m|n)$ Haldane--Shastry spin
  chain with the standard normalization (see, e.g., Ref.~\cite{BB06}).\qed
\end{remark}
It shall be useful for the sequel to define the sets
\newlength{\mylen} \settowidth{\mylen}{$\ds B_0=\{1,\dots,\lfloor(m+1)\rfloor/2\},$}
\begin{eqnarray*}
  &\makebox[\mylen][l]{$\ds B_{+}\!=\{1,\dots,\lfloor m/2\rfloor\},$}
    \qquad F_+=\{m+1,\dots,m+\lfloor n/2\rfloor\},\\
  &B_0=\{1,\dots,\lfloor(m+1)/2\rfloor\},\qquad F_0=\{m+1,\dots,m+\lfloor(n+1)/2\rfloor\},
\end{eqnarray*}
where $\lfloor x\rfloor$ denotes the integer part of the real number $x$. Note that $B_+=B_0$ for
$m$ even, while $B_0=B_+\cup\{(m+1)/2\}$ for $m$ odd, and similarly for $F_+$ and $F_0$. We shall
colloquially say that a spin $s_i$ is \emph{positive} (respectively\emph{ non-negative}), and
write $s_i\succ0$ (respectively $s_i\succeq0$) if $s_i\in B_+\cup F_+$ (respectively
$s_i\in B_0\cup F_0$).
\begin{remark}\label{rem.1}
  The operators $S_{ij}$ and $S_iS_j$ generate the Weyl algebra of the $D_N$ root system through
  the non-trivial relations
  \[
    S_{ij}^2=(S_iS_j)^2=1,\qquad S_{ij}S_{jk}=S_{ik}S_{ij}=S_{jk}S_{ik}, \qquad
    S_{ij}S_iS_k=S_jS_k S_{ij}\,,
  \]
  where the indices~$i,j,k$ are all distinct. Note, however, that the Hamiltonian~\eqref{Hchain}
  does \emph{not} have the standard form\footnote{From now on, unless otherwise stated all
    summations and products will range over the set~$\{1,\dots,N\}$.}
  \begin{equation}\label{cHDN}
    \cH_D=\sum_{i<j}\Big[f(\xi_i-\xi_j)(1-S_{ij})+f(\xi_i+\xi_j)(1-\tS_{ij})\Big]
  \end{equation}
  for a spin chain with \emph{real} sites~$\xi_k$ associated to the $D_N$ root
  system~\cite{BFG09,BFG11}. Nor is it purely of~$A_N$ type like, e.g., the supersymmetric version
  of the original HS chain~\cite{BB06}, due to the presence of the operators~$S_i$ in the
  Hamiltonian. In fact, the model~\eqref{Hchain} (as its non-supersymmetric version) is not
  directly associated to an extended root system, unlike most chains of HS type considered so far
  in the literature.\qed
\end{remark}

\begin{remark}
  The Hamiltonian~\eqref{Hchain} can be written as
  \begin{equation}\label{Hgen}
    \cH=\sum_{1\le i<j\le N}\left(\frac{1-S_{ij}}{|\ze_i-\ze_j|^2}
      +\frac{1-\tS_{ij}}{|\ze_i+\ze_j|^2}\right)
  \end{equation}
  in terms of the chain site coordinates $\ze_k=\e^{2\iu\th_k}$. The first term in the Hamiltonian
  is the usual spin-spin interaction between the spins at sites $\ze_i$ and $\ze_j$. On the other
  hand, the second term describes a non-standard interaction of the spin at site $\ze_i$ with the
  reflection with respect to the origin of the spin at site $\ze_j$. Note, finally, that the fact
  that the chain sites lie on the \emph{upper} unit half-circle means that, in spite of
  appearances, the chain~\eqref{Hchain} should be regarded as \emph{open}.\qed
\end{remark}

\medskip\noindent
An essential ingredient in the solvability of the chain~\eqref{Hchain} is its close connection
with the supersymmetric dynamical spin model with Hamiltonian
\begin{eqnarray}
  H&=-\De+8a\sum_{1\le i<j\le N}\bigg(\frac{a-S_{ij}}{|z_i-z_j|^2}
     +\frac{a-\tS_{ij}}{|z_i+z_j|^2}\bigg)\nonumber\\
   &=-\De+2a\sum_{i<j}\bigg[\frac{a-S_{ij}}{\sin^2(x_i-x_j)}
     +\frac{a-\tS_{ij}}{\cos^2(x_i-x_j)}\bigg]
     \label{Hspin}
\end{eqnarray}
and its scalar counterpart
\begin{eqnarray}
  \Hsc&=-\De+8a(a-1)\sum_{1\le i<j\le N}\bigg(\frac{1}{|z_i-z_j|^2}+\frac{1}{|z_i+z_j|^2}\bigg)
        \nonumber\\
      &=-\De+8a(a-1)\sum_{i<j}\sin^{-2}\Bigl(2(x_i-x_j)\Bigr),
        \label{Hsc}
\end{eqnarray}
where
\begin{equation}\label{zk}
  z_k=\e^{2\iu x_k},
\end{equation}
$\De:=\sum_i\frac{\pd^2}{\pd x_i^2}=-4\sum_i\left(z_i\frac{\pd}{\pd z_i}\right)^2$ and $a>1/2$.
Indeed, it can be readily checked that the chain sites~\eqref{roots} are the coordinates of the
minimum of the scalar potential
\[
  U(\bx)=\sum_{1\le i<j\le N}\bigg(\frac1{|z_i-z_j|^2} +\frac1{|z_i+z_j|^2}\bigg)=
  \sum_{i<j}\sin^{-2}\Bigl(2(x_i-x_j)\Bigr)
\]
of $\Hsc$ in the configuration space
\begin{equation}\label{Adef}
  A=\Big\{\bx:=(x_1,\dots,x_N)\in\RR^N:x_1<x_2\dots< x_N<x_1+\tfrac\pi2\Big\}\,,
\end{equation}
of $\Hsc$ and $H$ (this minimum is unique up to an inessential rigid translation~\cite{FG14}). We
obviously can write
\[
  H=\Hsc+8a\Hsp(\bx),
\]
where
\begin{equation}
  \Hsp(\bx):=\frac14\sum_{i<j}\bigg[\frac{1-S_{ij}}{\sin^2(x_i-x_j)}
  +\frac{1-\tS_{ij}}{\cos^2(x_i-x_j)}\bigg]\label{Hspinx}
\end{equation}
and
\[
  \Hsp(\th_1,\dots,\th_N)=\cH.
\]
Hence in the large coupling limit~$a\to\infty$ the energies of the spin dynamical
model~\eqref{Hspin} are approximately given by
\[
  E_{ij}\simeq E_i+8a\cE_j\,,
\]
where~$E_i$ and~$\cE_j$ are two arbitrary eigenvalues of the Hamiltonians~$\Hsc$ and~$\cH$,
respectively. The above relation, although not suited for computing the spectrum of $\cH$ in terms
of those of $H$ and $\Hsc$, yields however the following \emph{exact} formula for the partition
function~$\cZ$ of the chain~\eqref{Hchain}:
\begin{equation}
  \label{Zft}
  \cZ(T)=\lim_{a\to\infty}\frac{Z(8aT)}{\Zsc(8aT)}\,.
\end{equation}
\begin{remark}
  In view of Eq.~\eqref{zk}, it is natural to interpret the variables $x_k$ appearing in the
  Hamiltonians~\eqref{Hspin} and \eqref{Hsc} as half the \emph{angular} coordinates of the
  particles, and $z_k=\e^{2\iu x_k}$ as their actual (or \emph{physical}) coordinates in the unit
  circle. Thus the latter Hamiltonians describe the motion of a system of particles (with or
  without $\su(m|n)$ ``spin'') in the unit circle.\qed
\end{remark}
\begin{remark}
  The Hamiltonians $H$ and $H_{\text{sc}}$ in Eqs.~\eqref{Hspin}--\eqref{Hsc} are closely related
  to the Hamiltonians of the Sutherland spin and scalar dynamical models, respectively given by
  \[
    \fl
    H_{\text{S}}=-\De+2a\sum_{i<j}\sin^{-2}(x_i-x_j)(a-S_{ij}),\quad
    H_{\text{S,sc}}=-\De+2a(a-1)\sum_{i<j}\sin^{-2}(x_i-x_j).
  \]
  Indeed, if the supersymmetric spin reversal operators $S_i$ are replaced by plus or minus the
  identity then $H$ and $H_{\text{sc}}$ respectively reduce to $4H_{\text{S}}(2\bx)$ and
  $4H_{\text{S,sc}}(2\bx)$.\qed
\end{remark}
\section{Evaluation of the partition function}\label{sec.PF}

In this section we shall use the freezing trick formula~\eqref{Zft} to derive an exact expression
for the partition function of the supersymmetric chain~\eqref{Hchain}. To begin with, we quote the
formula derived in Ref.~\cite{BFG20} for the spectrum of the scalar Hamiltonian~\eqref{Hsc} in the
center of mass (CM) frame:
\begin{equation}\label{Ebn}
  E_{\bn}=4\sum_i\Big(2p_i+a(N-2i+1)\Big)^2,\qquad  p_i:= n_i-\frac{|\bn|}N.
\end{equation}
Here the multiindex $\bn:=(n_1,\dots,n_N)$ labeling the spectrum has non-negative integer
components $n_i$ satisfying
\[
  n_1\ge n_2\cdots\ge n_{N-1}\ge n_N=0,
\]
and we are using the notation
\[
  |\bn|:=\sum_in_i.
\]
Expanding $E_{\bn}$ in powers of~$a$ we find that
\begin{equation}\label{Enexp}
  E_{\bn}=E_0+16a\sum_ip_i(N-2i+1)+O(1),
\end{equation}
where
\begin{equation}\label{E0}
  E_0=4a^2\sum_i(N-2i+1)^2=\frac43\,N(N^2-1)a^2
\end{equation}
is the ground-state energy of~$\Hsc$ in the CM frame. Hence
\begin{equation}\label{Zsc}
  \fl
  \lim_{a\to\infty}q^{-\frac{E_0}{8a}}\Zsc(8aT)=\sum_{n_1\ge\cdots\ge n_{N-1}\ge0}q^{2\sum_ip_i(N+1-2i)}
  =\prod_{i=1}^{N-1}(1-q^{2i(N-i)})^{-1},
\end{equation}
where\footnote{For convenience, we are setting the Boltzmann constant $k_B$ equal to one.}
\[
  q:=\e^{-1/T}
\]
(see~Ref.~\cite{FG05} for details on the evaluation of the sum).

The computation of the spectrum of the dynamical spin Hamiltonian~$H$ in Eq.~\eqref{Hspin}
proceeds along same lines as for its non-supersymmetric version in Ref.~\cite{BFG20}. We shall
therefore omit unnecessary details, referring to the latter reference where needed. To begin with,
let us denote by $\La_s$ the total supersymmetric symmetrizer with respect to simultaneous
permutations of the particle's spatial coordinates and spins, determined by the relations
\[
  \Pi_{ij}\La_s=\La_s\Pi_{ij}=\La_s,\qquad 1\le i<j\le N.
\]
Here
\[
  \Pi_{ij}=K_{ij}S_{ij}=S_{ij}K_{ij},
\]
where $K_{ij}$ is the coordinate permutation operator defined by
\[
  K_{ij}f(z_1,\dots,z_i,\dots,z_j,\dots,z_N)=f(z_1,\dots,z_j,\dots,z_i,\dots,z_N),
\]
and $S_{ij}$ is the supersymmetric spin permutation operator acting on the internal (spin)
$\cS^{(m|n)}$ defined above. Note that $\La_s$ is an ordinary symmetrizer in the purely bosonic
case $n=0$, and an antisymmetrizer in the purely fermionic one $m=0$. For instance, for $N=2,3$ we
respectively have
\[
  \fl
  \La_{s}=\frac12(1+\Pi_{12}),\qquad
  \La_{s}=\frac16\,(1+\Pi_{12}+\Pi_{13}+\Pi_{23}+\Pi_{12}\Pi_{23}+\Pi_{12}\Pi_{13}).
\]
Likewise, we define the total (i.e., with respect to coordinates and spin variables)
supersymmetric flip operators $\Pi_i$ by
\[
  \Pi_i=K_iS_i=S_iK_i,
\]
where $K_i$ is the operator flipping the \emph{physical} coordinate of the $i$-th particle with
respect to the origin:
\[
  K_if(z_1,\dots,z_i,\dots,z_N)=f(z_1,\dots,-z_i,\dots,z_N).
\]
The projectors $\La_{0}^{\pm}$ onto the spaces of states even (``$+$'') or odd (``$-$'') under the
action of the supersymmetric flip operators $\Pi_i$ are then given by
\[
  \La_{0}^\vep=\frac1{N!}\left(1+\sum_{n=1}^N\sum_{i_1<\cdots<i_n}\vep^n \Pi_{i_1}\cdots
    \Pi_{i_n}\right),\qquad \vep=\pm1.
\]

Following the procedure in Ref.~\cite{BFG20}, we next construct a Schauder (i.e., non-orthonormal)
basis on which the dynamical spin Hamiltonian $H$ acts triangularly. To this end, we define the
scalar functions
\[
  \vp_{\bn}(\bz)=\mu(\bz)\prod_{i}z_i^{n_i},\qquad n_1\ge\cdots\ge n_N,
\]
where $n_i$ is an integer and
\[
  \mu(\bz):=\prod_{i<j}\abs{z_i-z_j}^a
\]
is the ground state wave function of the scalar Hamiltonian $\Hsc$. It can then be shown that the
states
\begin{equation}\label{Phibasis}
  \Phi^{\vep}_{\bn,\bbs}(\bz)=\bigg(\prod_iz_i\bigg)^{-\abs\bn/N}
  \La_s\La^\vep_0\left(\vp_{\bn}(\bz)\ket{\bbs}\right),\qquad\vep=\pm1,
\end{equation}
will in fact constitute a Schauder basis of the Hilbert space of $H$ provided that the quantum
numbers $\bn$ and $\bbs$ are chosen so that the set of all such states is linearly independent.
Note that the prefactor in the RHS of Eq.~\eqref{Phibasis} ensures that all the
states~$\Phi^{\vep}_{\bn,\bbs}$ have vanishing linear momentum, i.e., that we are working in the
CM frame. Moreover, since $\Phi^{\vep}_{\bn,\bbs}$ changes at most by a sign if we add to each
$n_i$ an arbitrary integer, we shall from now normalize the multiindex $\bn$ by taking $n_N=0$.
Following Ref.~\cite{BFG20}, we shall impose the following conditions on the quantum numbers
$\bn\in(\NN\cup\{0\})^N$ and $\bbs\in(B\cup F)^N$ to ensure the linear independence of the
states~\eqref{Phibasis}:
\begin{enumerate}[B1)]
\item $n_1\ge\cdots \ge n_N=0$.
\item $\ds n_i=n_j\en\implies\en s_i\ge s_j$, and $s_i>s_j$ if $s_i,s_j\in F$.
\item $s_i\in B_0\cup F_0$ if $(-1)^{n_i}\si(s_i)=\vep$, and $s_i\in B_+\cup F_+$ if
  $(-1)^{n_i}\si(s_i)=-\vep$.
\end{enumerate}
To understand these conditions, note that the states~\eqref{Phibasis} verify
\begin{eqnarray}
  \Phi^\vep_{P_{ij}\bn,P_{ij}\bbs}&=(-1)^{\nu(s_i,\dots,s_j)}\Phi^\vep_{\bn,\bbs},\label{Pij}\\
  \Phi^{\vep}_{\bn,P_i\bbs}&=\vep(-1)^{n_i}\si(s_i)\Phi^{\vep}_{\bn,\bbs},\label{Pi}
\end{eqnarray}
where $P_{ij}$ acts on the vectors $\bn$ and $\bbs$ by permuting their $i$-th and $j$-th
components and $P_i$ acts on the multiindex $\bbs$ by flipping its $i$-th component:
\[
  P_i(s_1,\dots,s_i,\dots,s_N)=(s_1,\dots,s_i',\dots,s_N).
\]
In particular, the wave functions $\Phi^\vep_{P_{ij}\bn,P_{ij}\bbs}$ and
$\Phi^{\vep}_{\bn,P_i\bbs}$ define the same quantum state as $\Phi^\vep_{\bn,\bbs}$. Thus, taking
advantage of~\eqref{Pij}, we can order the spin quantum numbers occupying the same positions as a
sequence of consecutive equal components of the multiindex $\bn$ by applying a suitable
permutation to $\bn$ and $\bbs$. This is the content of the second condition. As to the third one,
note first of all that if $s_i\not\in B_0\cup F_0$ then $s_i'\in B_0\cup F_0$, and by
Eq.~\eqref{Pi} the spin flip $s_i\mapsto s_i'$ does not change the state. Thus we can take
$s_i\in B_0\cup F_0$ by flipping the $i$-th spin if necessary. Whether $s_i$ can be equal to
$(m+1)/2$ (when $m$ is odd) or $m+1+(n+1)/2$ (when $n$ is odd) depends on the value of $\vep$.
Indeed, in either case $P_i\bbs=\bbs$, and hence~\eqref{Pi} implies that
\[
  \Phi^\vep_{\bn,\bbs}=\Phi^\vep_{\bn,P_i\bbs}=\vep(-1)^{n_i}\si(s_i)\Phi^\vep_{\bn,\bbs}
  \en\implies\en (-1)^{n_i}\si(s_i)=\vep.
\]
Thus $s_i$ can take the value $(m+1)/2$ (when $m$ is odd) or $m+1+(n+1)/2$ (when $n$ is odd) if
and only if $(-1)^{n_i}\si(s_i)=\vep$. The last two remarks thus account for condition~iii).

Proceeding in much the same way as for the non-supersymmetric case studied in Ref.~\cite{BFG20},
one can show that the states~\eqref{Phibasis} with quantum numbers $\bn$ and $\bbs$ obeying
conditions~B1)--B3) above can be ordered in such a way that the action of the spin Hamiltonian $H$
on them is triangular. It follows that the eigenvalues of $H$ (in the CM frame) coincide with its
diagonal matrix elements in the latter basis, given by
\begin{equation}
  \label{En}
  E^{\vep}_{\bn,\bbs}=4\sum_i\Big(p_i+a(N+1-2i)\Big)^2=E_\bn,\qquad p_i=n_i-\frac{\abs{\bn}}N,
\end{equation}
where the quantum numbers $\vep=\pm1$, $\bn$, $\bbs$ satisfy conditions B1)--B3) (see
Ref~\cite{BFG20} for the details). Since the RHS of this equation depends only on the multiindex
$\bn$, the partition function of the dynamical spin Hamiltonian~\eqref{Hspin} is given by
\begin{equation}
  \label{PFin}
  Z(q)=\sum_{n_1\ge\dots\ge n_{N-1}\ge0}D(\bn)\,q^{E_\bn},
\end{equation}
where the \emph{spin degeneracy} $D(\bn)$ is equal to the number of choices of pairs $(\vep,\bbs)$
satisfying conditions B1)--B3) for a given multiindex $\bn$ (with $n_1\ge\cdots\ge n_N=0$). Using
the expansion~\eqref{Enexp}-\eqref{E0} of $E_\bn$ in powers of $a$ we obtain
\begin{equation}\label{Zspin}
  \lim_{a\to\infty}q^{-\frac{E_0}{8a}}Z(8aT)=\sum_{n_1\ge\cdots\ge n_{N-1}\ge0}D(\bn)\,q^{\sum_i p_i(N+1-2i)}.
\end{equation}
From Eqs.~\eqref{Zft}, \eqref{Zsc}, and \eqref{Zspin} we then deduce the following formula for the
partition function of the supersymmetric chain~\eqref{Hchain}:
\begin{equation}\label{Zchainin}
  \cZ(q)=\prod_{i=1}^{N-1}\left(1-q^{2i(N-i)}\right)\cdot
  \sum_{n_1\ge\cdots\ge n_{N-1}\ge0}D(\bn)\,q^{\sum_i p_i(N+1-2i)}.
\end{equation}
The first step in the evaluation of the partition function $\cZ(q)$ is thus a combinatorial
problem, namely the evaluation of the spin degeneracy $D(\bn)$ for an arbitrary integer multiindex
$\bn=(n_1,\dots,n_N)$ satisfying $n_1\ge\cdots\ge n_N=0$. Due to condition~B3), it is clear that
this combinatorial problem depends crucially on the parity of $m$ and $n$. Before starting a
detailed analysis, it is convenient to parametrize the multiindex $\bn$ as follows:
\begin{equation}\label{bnnu}
  \bn=\big(\underbrace{\nu_1,\dots,\nu_1}_{\ell_1}\,,\dots,
  \underbrace{\nu_{r-1},\dots,\nu_{r-1}}_{\ell_{r-1}}\,,\underbrace{0,\dots,0}_{\ell_r}\big)\,,
\end{equation}
where $\nu_1>\cdots>\nu_{r-1}>\nu_r=0$ (with~$\nu_k\in\ZZ$), $\ell_1+\cdots +\ell_r=N$
(with~$\ell_i>0$ for all $i$) and $r=1,\dots,N$. Conditions B1)--B3) imply that
\begin{equation}\label{Dbn}
  D(\bn)=\prod_{i=1}^rd_+(\ell_i,\nu_i)+\prod_{i=1}^rd_-(\ell_i,\nu_i)=:d(\bell,\bnu),
\end{equation}
where $\bell=(\ell_1,\dots,\ell_r)$, $\bnu=(\nu_1,\dots,\nu_{r-1},\nu_r=0)$, and
$d_\vep(\ell,\nu)$ is the number of choices of $\ell$ spins $\{s_1,\dots,s_\ell\}$ satisfying
\begin{enumerate}[C1)]
\item $s_i\ge s_j$, and $s_i>s_j$ if $s_i,s_j\in F$.
\item $s_i\in B_0\cup F_0$ if $(-1)^{\nu}\si(s_i)=\vep$, and $s_i\in B_+\cup F_+$ if
  $(-1)^{\nu}\si(s_i)=-\vep$.
\end{enumerate}
Thus $d_\pm(\ell,\nu)$ is the contribution to the spin degeneracy of a constant ``sector''
$(\nu,\dots,\nu)$ of length $\ell$ in the multiindex $\bn$ when $\vep$ takes the fixed value
$\pm1$. It is also clear that
\begin{equation}\label{dellnu}
  d_\pm(\ell,\nu)=\sum_{k=0}^{\ell}d_\pm^B(k,\nu)d_\pm^F(\ell-k,\nu),
\end{equation}
where $d_\pm^B(k,\nu)$ denotes the number of choices of $k$ bosonic spin satisfying
conditions~C1)--C2) when $\vep=\pm1$, and similarly for $d_\pm^F(\ell-k,\nu)$. From the
parametrization~\eqref{bnnu} of the multiindex $\bn$ and Eqs.~\eqref{Zchainin}-\eqref{Dbn} we
obtain the following formula for the partition function of the supersymmetric chain~\eqref{Hchain}
\begin{equation}
  \label{Zdlnu}
  \fl
  \cZ(q)=\prod_{i=1}^{N-1}\left(1-q^{2i(N-i)}\right)
  \sum_{r=1}^N\sum_{\bell\in\cP_N(r)}\sum_{\nu_1>\cdots>\nu_{r-1}>0}d(\bell,\bnu)\,q^{\sum_i p_i(N+1-2i)},
\end{equation}
where $\cP_N(r)$ denotes the set of compositions (i.e., ordered partitions) of the integer $N$
into $r$ parts.

\subsection{$m$ and $n$ even}\label{sec.ee}

In this case condition C2) simplifies to
\begin{enumerate}[C2$'$)]
\item $s_i\in B_+\cup F_+$,
\end{enumerate}
independently of the value of $\vep$, $\nu$ and $\vep_F$. In particular, in this case the
partition function is the same for $\vep_F=1$ and $\vep_F=-1$. Since the cardinals of the sets
$B_+$ and $F_+$ are respectively $m/2$ and $n/2$, by condition C1) we have\footnote{Here and in
  what follows we shall set $\binom{n}k=0$ for $k>n\ge0$ and $\binom{n}0=1$ for $n\le0$.}

\begin{equation}\label{dpmBF}
  d_\pm^B(k,\nu)=\binom{\frac{m}2+k-1}{k},\qquad d_\pm^F(\ell-k,\nu)=\binom{\frac{n}2}{\ell-k},
\end{equation}
and therefore
\begin{equation}\label{dlnuee}
  d_\pm(\ell,\nu)=\sum_{k=0}^\ell\binom{\frac{m}2+k-1}{k}\binom{\frac{n}2}{\ell-k}.
\end{equation}
Thus $d_\vep(\ell,\nu)\equiv d(\ell)$ is in this case independent of $\nu$ and $\vep$. It follows
that
\[
  d(\bell,\bnu)=2\prod_{k=1}^rd(\ell_k)\equiv d(\bell),
\]
where the factor of $2$ takes care of the two possible choices of $\vep=\pm1$, and
Eq.~\eqref{Zdlnu} simplifies to
\begin{equation}\label{cZsum}
  \fl
  \cZ(q)=2\prod_{i=1}^{N-1}\left(1-q^{2i(N-i)}\right)
  \sum_{r=1}^N\sum_{\bell\in\cP_N(r)}\prod_{k=1}^rd(\ell_k)\sum_{\nu_1>\cdots>\nu_{r-1}>0}q^{\sum_i
    p_i(N+1-2i)}.
\end{equation}
To evaluate the last sum we define
\[
  \tnu_i=\nu_i-\nu_{i+1},\qquad i=1,\dots,r-1,
\]
with $\nu_r=0$, so that the sum over $\nu_1>\cdots>\nu_{r-1}>0$ turns into an unrestricted sum
over the positive integers $\tnu_1,\dots,\tnu_{r-1}$. Taking into account that
\[
  \sum_ip_i(N+1-2i)=\sum_{i=1}^{r-1}\tnu_i\cE(L_i)\,,\qquad L_i:=\sum_{j=1}^i\ell_j\,,
\]
where the \emph{dispersion relation} $\cE$ is given by
\[
  \cE(i)=i(N-i)
\]
(cf.~\cite{FG05,BFG20}), we readily obtain
\[
  \fl \sum_{\nu_1>\cdots>
    \nu_{r-1}>0}q^{\sum_ip_i(N+1-2i)}=\sum_{\tnu_1,\dots,\tnu_{r-1}=1}^\infty
  \prod_{i=1}^{r-1}q^{\tnu_i\cE(L_i)} =\prod_{i=1}^{r-1}\sum_{\tnu_i=1}^\infty q^{\tnu_i\cE(L_i)}
  =\prod_{i=1}^{r-1}\frac{q^{\cE(L_i)}}{1-q^{\cE(L_i)}}\,.
\]
Substituting into Eq.~\eqref{cZsum} we arrive at the following explicit function for the partition
function of the supersymmetric chain~\eqref{Hchain} in this case:
\begin{equation}
  \label{cZee}
  \fl
  \cZ(q)=
  2\prod_{i=1}^{N-1}\Big(1+q^{\cE(i)}\Big)
  \cdot\sum_{r=1}^N\sum_{\bell\in\cP_N(r)} \prod_{i=1}^rd(\ell_i)\cdot
  q^{\sum\limits_{i=1}^{r-1}\cE(L_i)}
  \prod_{i=1}^{N-r}\Big(1-q^{\cE(L_i')}\Big)\,,
\end{equation}
with $d(\ell)$ defined in Eq.~\eqref{dlnuee}
\[
  \{L_1',\dots,L_{N-r}'\}=\{1,\dots,N-1\}\setminus\{L_1,\dots,L_{r-1}\}.
\]
Thus when both $m$ and $n$ are even the partition function of the $\su(m|n)$ supersymmetric
chain~\eqref{Hchain} can be factored as the product
\begin{equation}\label{fact}
  \cZ^{(m|n)}(q)=\cZ_{\mathrm{HS}}^{(1|1)}(q)\,\cZ_{\mathrm{HS}}^{\bigl(\frac m2|\frac n2\bigr)}(q),
\end{equation}
of the partition functions of the $\su(1|1)$ and~$\su\bigl(\tfrac m2|\tfrac n2\bigr)$ HS spin
chains~\cite{BB06}. In other words, in this case the supersymmetric chain
Hamiltonian~\eqref{Hchain} is unitarily equivalent to the sum
\begin{equation}\label{HHS}
  \cH_{\mathrm{HS}}^{(1|1)}\otimes\id+\id\otimes\cH_{\mathrm{HS}}^{\bigl(\frac m2|\frac n2\bigr)},
\end{equation}
where $\cH_{\mathrm{HS}}^{(p|q)}$ denotes the Hamiltonian of the $\su(p|q)$ HS chain. This
property, which is not evident from Eq.~\eqref{Hchain}, is the essential ingredient used to arrive
at the description of the model's spectrum in terms of Haldane motifs and their corresponding
Young tableaux akin to the one developed in Ref.~\cite{BFG20} for the non-supersymmetric case.

The combinatorial formula~\eqref{cZee} becomes particularly simple when $m,n\le 2$. Indeed, when
$m=2$ and $n=0$ by Eq.~\eqref{dlnuee} we have
\[
  \prod_{i=1}^rd(\ell_i)=\prod_{i=1}^r\binom{i}{i}=1,
\]
so that
\[
  \cZ^{(1|0)}_{\mathrm{HS}}(q)=\sum_{r=1}^N\sum_{\bell\in\cP_N(r)}
  q^{\sum\limits_{i=1}^{r-1}\cE(L_i)}\,\prod_{i=1}^{N-r}\Big(1-q^{\cE(L_i')}\Big)
  =\prod_{i=1}^N\left(1-q^{\cE(i)}+q^{\cE(i)}\right)=1,
\]
and therefore
\begin{equation}\label{cZ20}
  \cZ^{(2|0)}(q)=\cZ^{(1|1)}_{\mathrm{HS}}(q)=2\prod_{i=1}^{N-1}\Big(1+q^{\cE(i)}\Big).
\end{equation}
Similarly, when $m=0$ and $n=2$ the spin degeneracy is given by
\[
  \prod_{i=1}^rd(\ell_i)=\cases{1,& $\ell_i=1\en\forall i$,\\ 0,&otherwise,}
\]
so that
\[
  \cZ^{(0|1)}_{\mathrm{HS}}(q)=q^{\sum_{i=1}^{N-1}\cE(i)}=q^{\frac16N(N^2-1)}
\]
and
\begin{equation}\label{cZ02}
  \fl
  \cZ^{(0|2)}(q)=q^{\frac16N(N^2-1)}\cZ^{(1|1)}_{\mathrm{HS}}(q)=
  2q^{\frac16N(N^2-1)}\prod_{i=1}^{N-1}\Big(1+q^{\cE(i)}\Big)
  =q^{\frac16N(N^2-1)}\cZ^{(2|0)}(q).
\end{equation}
Finally, when $m=n=2$ we have
\[
  \cZ^{(2|2)}(q)=\left[\cZ_{\mathrm{HS}}^{(1|1)}(q)\right]^2=4\prod_{i=1}^{N-1}\Big(1+q^{\cE(i)}\Big)^2;
\]
in particular, in this case the energy levels are at least four times degenerate.

In general, from Eq.~\eqref{cZee} it follows that in this case all the energy levels have even
degeneracy. This is an immediate consequence of the independence of the spin degeneracy factor
from the value of $\vep=\pm1$. Another direct consequence of Eq.~\eqref{cZee} is the fact that,
with the normalization chosen for the Hamiltonian~\eqref{Hchain}, all energies are nonnegative
integers. Furthermore, the degeneracy of the model's zero mode is given by
\[
  \cZ(0)=2\sum_{r=1}^N\sum_{\bell\in\cP_N(r)} \prod_{i=1}^rd(\ell_i)\cdot
  q^{\sum\limits_{i=1}^{r-1}\cE(L_i)}\bigg|_{q=0}.
\]
A cursory inspection of the latter formula shows that the only partition of $N$ contributing to
the zero mode is $\bell=(N)$, so that
\begin{equation}\label{dGSee}
  \cZ(0)=d((N))=2\sum_{k=0}^N\binom{\frac{m}2+k-1}{k}\binom{\frac{n}2}{N-k}.
\end{equation}
Since the RHS does not vanish unless $m=0$, we conclude that the ground state degeneracy of the
supersymmetric chain~\eqref{Hchain} with $m\ne0$ and $n$ both even coincides with the RHS of
Eq.~\eqref{dGSee}, which is in turn twice the ground state degeneracy of the
$\su\big(\tfrac{m}2,\tfrac{n}2\big)$ HS chain.

\subsection{$m$ even and $n$ odd}

In this case condition~C2) simply states that $s_i\in B_+$ if $s_i$ is bosonic, and therefore
$d^B_\pm(k,\nu)$ is still given by Eq.~\eqref{dpmBF}. On the other hand, for fermionic spins
condition C2) reads
\begin{enumerate}[C2F)]
\item $s_i\in F_0$ if $(-1)^{\nu}\vep_F=\vep$, and $s_i\in F_+$ if $(-1)^{\nu}\vep_F=-\vep$.
\end{enumerate}
Since
\[
  \Abs{F_0}=\frac{n+1}2,\qquad \Abs{F_+}=\frac{n-1}2,
\]
a moment's thought reveals that
\begin{equation}\label{dFodd}
  d_\pm^F(\ell-k,\nu)=\binom{\frac12(n\pm\vep_F(-1)^\nu)}{\ell-k},
\end{equation}
and therefore
\begin{equation}
  \label{dlnueo}
  d_\pm(\ell,\nu)=\sum_{k=0}^\ell\binom{\frac{m}2+k-1}{k}
  \binom{\frac12(n\pm\vep_F(-1)^\nu)}{\ell-k}.
\end{equation}
Hence
\[
  d_\pm(\ell,\nu;\vep_F=1)=d_\mp(\ell,\nu;\vep_F=-1),
\]
from which it follows that the spin degeneracy $d(\bell,\bnu)$ given by Eq.~\eqref{Dbn} ---and
hence the partition function~\eqref{Zdlnu}--- is again independent of $\vep_F$. As in the previous
case, it is convenient to rewrite $\cZ(q)$ in terms of the independent variables
$\tnu_i=\nu_i-\nu_{i+1}$, namely
\begin{equation}\label{cZsumeo}
  \fl
  \cZ(q)=\prod_{i=1}^{N-1}\left(1-q^{2\cE(i)}\right)
  \sum_{r=1}^N\sum_{\bell\in\cP_N(r)}\sum_{\tnu_1,\dots,\tnu_{r-1}=1}^\infty
  d(\bell,\bnu)\,q^{\sum_{i=1}^{r-1}\tnu_i\cE(L_i)},
\end{equation}
with
\[
  \nu_i=\sum_{j=i}^{r-1}\tnu_j.
\]
Since in this case $d_\pm(\ell,\nu)$ depends only on the parity of the integer $\nu$, it is
convenient to define
\[
  \tnu_i=2\tn_i-\de_i,
\]
where $\tn_i\in\NN$ and $\de_i\in\{0,1\}$. We then have
\begin{equation}\label{Dei}
  (-1)^{\nu_i}=(-1)^{\De_i},\qquad\text{with}\quad \De_i:=\sum_{k=i}^{r-1}\de_k,
\end{equation}
and therefore
\begin{eqnarray}
  \fl
  \sum_{\tnu_1,\dots,\tnu_{r-1}=1}^\infty d(\bell,\bnu)\,
  q^{\sum_{i=1}^{r-1}\tnu_i\cE(L_i)}
  &= \sum_{\de_1,\dots,\de_{r-1}=0}^1  d(\bell,\bDe)\sum_{\tn_1,\dots,\tn_{r-1}=1}^\infty
    q^{\sum_{i=1}^{r-1}(2\tn_i-\de_i)\cE(L_i)}\nonumber\\
  \fl
  &=\sum_{\de_1,\dots,\de_{r-1}=0}^1d(\bell,\bDe)
    \prod_{i=1}^{r-1}\frac{q^{(2-\de_i)\cE(L_i)}}{1-q^{2\cE(L_i)}},
    \label{sumdlnu}
\end{eqnarray}
where $\bDe=(\De_1,\dots,\De_{r-1},\De_r=0)$. Combining this equation with Eq.~\eqref{cZsumeo} we
finally obtain the following explicit formula for the partition function $\cZ(q)$:
\begin{equation}
  \label{cZeo}
  \fl
  \cZ(q)=
  \sum_{r=1}^N\sum_{\bell\in\cP_N(r)}\prod_{i=1}^{N-r}\Big(1-q^{2\cE(L_i')}\Big)
  \sum_{\de_1,\dots,\de_{r-1}=0}^1d(\bell,\bDe)\,
  q^{\sum_{i=1}^{r-1}(2-\de_i)\cE(L_i)}.
\end{equation}

The spin degeneracy factor can be somewhat simplified when $m=2$ and $n=1$. Indeed, for $n=1$ the
factor
\[
  \binom{\frac12(n\pm\vep_F(-1)^\nu)}{\ell-k}=\binom{\frac12(1\pm\vep_F(-1)^\nu)}{\ell-k}
\]
in Eq.~\eqref{dlnueo} is equal to $1$ if $(-1)^\nu=\mp\vep_F$ and $k=\ell$, or
$(-1)^\nu=\pm\vep_F$ and $k=\ell-1,\ell$, and is zero otherwise. Thus for $m=2$ and $n=1$
Eq.~\eqref{dlnueo} reduces to
\[
  d_\pm(\ell,\nu)=1+\frac12(1\pm\vep_F(-1)^\nu),
\]
whence it follows that
\begin{eqnarray*}
  d_\pm(\bell,\bnu)=&2^{\Abs{\big\{i:(-1)^{\nu_i}=\pm\vep_F,\ 1\le i\le r\big\}}}\\
                    &\implies\en
                      d(\bell,\bnu)=2^{\Abs{\big\{i:(-1)^{\nu_i}=-1,\ 1\le i\le
                      r\big\}}}+2^{\Abs{\big\{i:(-1)^{\nu_i}=1,\ 1\le i\le r\big\}}},
\end{eqnarray*}
where $\abs A$ denotes the cardinal of the set $A$.

As in the previous case, from Eq.~\eqref{cZeo} it follows that the energies are nonnegative
integers, and by Eqs.~\eqref{Dbn} and \eqref{dlnueo} the degeneracy of the zero mode is given by
\begin{equation}\label{GSdegEO}
  \fl
  \cZ(0)=d((N),(0))=\sum_{k=0}^N\binom{\frac{m}2+k-1}{k}\left[\binom{\frac12(n-1)}{N-k}
    +\binom{\frac12(n+1)}{N-k}\right].
\end{equation}
Since the right hand side does not vanish for $m\ne0$, it follows that when $m\ne0$ the ground
state is the zero mode, with degeneracy equal to the RHS of the latter equation. In particular,
for $n=1$ and arbitrary (even) $m$ the ground state degeneracy is simply
\[
  2\binom{\frac m2+N-1}{N}+\binom{\frac m2+N-2}{N-1}.
\]
Thus for $m=2$ and $n=1$ the ground state of the supersymmetric chain~\eqref{Hchain} has zero
energy and is three times degenerate, regardless of the value of $N$.

\subsection{$m$ odd and $n$ even}

This case is similar to the previous one, with the roles of $m$ and $n$ reversed. In other words,
\begin{equation}\label{dBFeo}
  d_\pm^B(k,\nu)=\binom{\frac12(m\pm(-1)^\nu)+k-1}{k},\qquad
  d_{\pm}^F(\ell-k,\nu)=\binom{\frac{n}2}{\ell-k}
\end{equation}
for both $\vep_F=1$ and $\vep_F=-1$, since as $n$ is even condition C3) for fermions simply states
that $s_i\in F_+$. We thus have
\begin{equation}
  d_\pm(\ell,\nu)=\sum_{k=0}^\ell\binom{\frac{n}2}{\ell-k}\binom{\frac12(m\pm(-1)^\nu)+k-1}{k}.
  \label{dlnuoe}
\end{equation}
Equation~\eqref{cZeo} for the partition function is still valid in this case ---with
$d_\pm(\ell,\nu)$ defined by Eq.~\eqref{dlnuoe} instead of~\eqref{dlnueo}---, since in its
derivation we only used the fact that $d(\ell,\nu)$ in Eq.~\eqref{dlnueo} depended on $\nu$
through $(-1)^\nu$.

As in the previous case, the spin degeneracy factor simplifies to some extent when $m=1$. Indeed,
in this case the second binomial coefficient in Eq.~\eqref{dlnuoe} is equal to $1$ for all values
of $k=0,\dots,\ell$ when $(-1)^\nu=\vep$, so that
\[
  d_\vep(\ell,\nu)=\sum_{k=0}^\ell\binom{\frac{n}2}{\ell-k},\qquad (-1)^\nu=\vep.
\]
In particular,
\[
  d_\vep(\ell,\nu)=2^{n/2},\qquad (-1)^\nu=\vep\quad\text{and}\quad l\ge \frac n2.
\]
On the other hand, when $(-1)^\nu=-\vep$ the latter binomial coefficient reduces to
\[
  \binom{k-1}{k}=\cases{1,& $k=0$\\ 0, & $1\le k\le\ell$,}
\]
and therefore
\[
  d_\vep(\ell,\nu)=\binom{\frac{n}2}{\ell},\qquad (-1)^\nu=-\vep.
\]

As in the previous cases, the energies are nonnegative integers, and the degeneracy of the zero
mode can be easily computed from the formula
\begin{equation}\label{dGSoe}
  \fl
  \cZ(0)=d((N),(0))=\sum_{k=0}^N\binom{\frac{n}2}{N-k}\left[\binom{\frac12(m-1)+k-1}k
    +\binom{\frac12(m+1)+k-1}k\right].
\end{equation}
Since this number does not vanish, it follows that the ground state has again energy zero and
degeneracy given by the previous equation. In particular, for $m=1$ the ground state degeneracy is
\[
  \binom{\frac{n}2}{N}+\sum_{k=0}^N\binom{\frac n2}k,
\]
which reduces to $2^{n/2}$ (independent of $N$) when $N>n/2$.

\subsection{$m$ and $n$ odd}

In this case the fermionic spins satisfy condition C2F) above, so that $d_\pm^F(\ell-k,\nu)$ is
given by Eq.~\eqref{dFodd}. The bosonic spins verify the analogous condition
\begin{enumerate}[C2B)]
\item $s_i\in B_0$ if $(-1)^{\nu}=\vep$, and $s_i\in B_+$ if $(-1)^{\nu}=-\vep$,
\end{enumerate}
so that $d_\pm^B(k,\nu)$ is given by Eq.~\eqref{dlnuoe}. From Eq.~\eqref{dellnu} it then follows
that
\begin{equation}
  d_\pm(\ell,\nu;\vep_F)
  =\sum_{k=0}^\ell
  \binom{\frac12(m\pm(-1)^\nu)+k-1}{k}\binom{\frac12(n\pm\vep_F(-1)^\nu)}{\ell-k}.
  \label{dlnuoo}
\end{equation}
Note that in this case $d_\pm(\ell,\nu;\vep_F=1)\ne d_\mp(\ell,\nu;\vep_F=-1)$, and hence the
degeneracy factor $d(\bell,\bnu)$ in Eq.~\eqref{Dbn} is \emph{not} the same for both values of
$\vep_F$. More precisely,
\begin{eqnarray}
  \fl
  d(\bell,\bnu;\vep_F=1)
  &=\prod_{i=1}^r\sum_{k=0}^{\ell_i}
    \binom{\frac12(m+(-1)^{\nu_i})+k-1}{k}\binom{\frac12(n+(-1)^{\nu_i})}{\ell_i-k}\nonumber\\
  &\quad
    +\prod_{i=1}^r\sum_{k=0}^{\ell_i}
    \binom{\frac12(m-(-1)^{\nu_i})+k-1}{k}\binom{\frac12(n-(-1)^{\nu_i})}{\ell_i-k},
    \label{dlnu1}\\
  \fl
  d(\bell,\bnu;\vep_F=-1)
  &=\prod_{i=1}^r\sum_{k=0}^{\ell_i}
    \binom{\frac12(m+(-1)^{\nu_i})+k-1}{k}\binom{\frac12(n-(-1)^{\nu_i})}{\ell_i-k}\nonumber\\
  &\quad
    +\prod_{i=1}^r\sum_{k=0}^{\ell_i}
    \binom{\frac12(m-(-1)^{\nu_i})+k-1}{k}\binom{\frac12(n+(-1)^{\nu_i})}{\ell_i-k}.
    \label{dlnu2}
\end{eqnarray}
The partition function in this case is again given by Eq.~\eqref{cZeo}, with $d(\bell,\bnu)$
replaced by $d(\bell,\bnu;\vep_F)$ in Eqs.~\eqref{dlnu1}-\eqref{dlnu2}. In particular,
\[
  \cZ(q;\vep_F=1)\ne\cZ(q;\vep_F=-1).
\]

As an example, consider the simplest case $m=n=1$. To begin with, if $\vep_F=1$ the spin flip
operator $S_i$ reduces to the identity. Hence $\tS_{ij}=S_{ij}\equiv S_{ij}^{(1|1)}$ and
\[
  \fl \cH=\frac14\sum_{i<j}(1-S_{ij}^{(1|1)})\left(\sin^{-2}(\th_i-\th_j)
    +\cos^{-2}(\th_i-\th_j)\right)
  =\sum_{i<j}\frac{1-S_{ij}^{(1|1)}}{\sin^2\bigl(2(\th_i-\th_j)\bigr)}\,,
\]
which (apart from a conventional factor of $1/2$) is the Hamiltonian of the $\su(1|1)$
Haldane--Shastry model~\cite{BB06}.

Consider next the case $m=n=1$ and $\vep_F=-1$. Much as in the previous cases, the chain's
energies in this case are nonnegative integers, and the zero mode degeneracy can be readily
computed from Eq.~\eqref{cZeo} for the partition function from the usual formula
$\cZ(0)=d((N),(0))$. In general, when $m=n=1$ we have
\begin{eqnarray*}
  \binom{\frac12(1+\vep(-1)^\nu)+k-1}{k}&=\cases{1,&$(-1)^\nu=\vep$\\
  1,&$(-1)^\nu=-\vep$ and $k=0$\\
  0,&$(-1)^\nu=-\vep$ and $k>0$}\\
  \binom{\frac12(1-\vep(-1)^\nu)}{\ell-k}&=\cases{1,&$(-1)^\nu=\vep$ and $k=\ell$\\
  1,&$(-1)^\nu=-\vep$ and $k=\ell-1,\ell$\\
  0,&otherwise,}
\end{eqnarray*}
and therefore (assuming, again, that $\vep_F=-1$)
\[
  d_\vep(\ell,\nu)=\cases{1,&$(-1)^\nu=\vep$\\ 1, &$(-1)^{\nu}=-\vep$ and $\ell=1$\\
    0,&$(-1)^{\nu}=-\vep$ and $\ell>1$. }
\]
It follows that $\prod_{i=1}^rd_\vep(\ell_i,\nu_i)$ is equal to $1$ if $\ell_i=1$ for all $i$ such
that $(-1)^{\nu_i}=-\vep$, and vanishes otherwise. We thus see that in this case the spin
degeneracy depends crucially on the number of components of $\bell$ greater than $1$, or
equivalently of sectors of length greater than one in the multiindex~$\bn$. More precisely,
calling
\[
  \la(\bell)=\Abs{\big\{i:\ell_i>1,\ 1\le i\le r\big\}},
\]
the spin degeneracy of the $\su(1,1)$ supersymmetric chain~\eqref{Hchain} is given by
\begin{equation}
  \label{spind11}
  d(\bell,\bnu)=\cases{2,&$\bell=(1,\dots,1)$\\
    1,& $\la(\bell)=1$\\
    1,& $\la(\bell)>1$ and $(-1)^{\nu_i}=(-1)^{\nu_j}$ if $\ell_i,\ell_j>1$\\
    0,& otherwise.}
\end{equation}
In particular, the degeneracy of the zero mode is in this case
\[
  \cZ(0)=d((N),(0))=1.
\]
In other words, when $m=n=1$ and $\vep_F=-1$ the ground state is non-degenerate for all values of
$N$. Combining Eq.~\eqref{spind11} for the spin degeneracy with Eqs.~\eqref{cZsumeo} and
\eqref{sumdlnu} we also obtain the following more explicit formula for the partition function of
the $\su(1|1)$ chain:
\begin{eqnarray*}
  \fl
  \cZ^{(1|1)}(q;\vep_F=-1)=
  &q^{e(N)}\cZ^{(1|1)}_{\mathrm{HS}}(q)\\
  \fl
  &
    +\prod_{i=1}^{N-1}(1+q^{\cE(i)})\sum_{1\le k<j\le N}
    q^{e(k)+e(N-j+1)}
    \prod_{i=k}^{j-1}(1-q^{\cE(i)})\\
  \fl
  &+\sum_{r=2}^{N-2}\sum_{\substack{\bell\in\cP_N(r)\cr\la(\bell)>1}}
    \prod_{i=1}^{N-r}\Big(1-q^{2\cE(L_i')}\Big)
    \sum_{\substack{\de_1,\dots,\de_{r-1}\in\{0,1\}\cr
    \ell_i,\ell_j>1\:\Rightarrow\:(-1)^{\De_i}=(-1)^{\De_{\smash j}}}}
    q^{\sum_{i=1}^{r-1}(2-\de_i)\cE(L_i)},
\end{eqnarray*}
where
\[
  e(k)=\sum_{i=1}^{k-1}\cE(i)=\frac16k(k-1)(3N-2k+1).
\]
\label{ejk}

Proceeding as above it is straightforward to show that for both $\vep_F=1$ and $\vep_F=-1$ the
ground state is the zero mode, and its degeneracy is given by
\begin{eqnarray}
  \fl
  \cZ(0)=d((N),(0))=\sum_{k=0}^N
  &\left[\binom{\frac12(m+1)+k-1}{k}\binom{\frac12(n+\vep_F)}{N-k}\right.\nonumber\\
  &\hphantom{\frac12(m+1)+k}\left.+\binom{\frac12(m-1)+k-1}{k}\binom{\frac12(n-\vep_F)}{N-k}\right]
  \label{GSdegOO}
\end{eqnarray}
In particular,
\[
  \fl \cZ(0)=\cases{\binom{\frac12(m-1)+N}N+\binom{\frac12(m-1)+N-1}N
    +\binom{\frac12(m+\vep_F)+N-2}{N-1},&$n=1$\\
    2^{\frac12(n+\vep_F)}+\binom{\frac12(n-\vep_F)}N,& $m=1$.}
\]

\section{Symmetries}\label{sec.symm}

In this Section we shall examine two basic symmetries of the supersymmetric chain~\eqref{Hchain},
namely invariance under ``twisted'' translations (defined below) and boson-fermion duality.

\subsection{Twisted translations}

Although the interaction strengths in Eq.~\eqref{Hchain} depend on the differences $\th_i-\th_j$,
the model clearly lacks translation invariance due to its fundamentally \emph{open} nature.
Indeed, while a formal ``translation'' $k\mapsto k+1$ along the chain maps the $k$-th site located
at $\ze_k=\e^{2\iu\th_k}$ into the $(k+1)$-th for $1\le k<N$, the $N$-th site at
$\ze_N=\e^{2\iu\th_N}=-1$ is mapped to the point $\ze_{N+1}=-\e^{\iu\pi/N}\ne\ze_1=\e^{\iu\pi/N}$.
The lack of translation invariance of the Hamiltonian~\eqref{Hchain} can be more formally shown by
computing the action on the $\cH$ of the (left) translation operator $T$, defined by the relation
\[
  T\ket{s_1,\dots,s_N}:=\ket{s_2,\dots,s_N,s_1}\,.
\]
Clearly $T^\dagger=T^{-1}$ and
\[
  T^\dagger S_{ij}T=S_{i+1,j+1},\qquad T^\dagger \tS_{ij}T=\tS_{i+1,j+1}
\]
with
\begin{equation}\label{percond}
  S_{k,N+1}\equiv S_{1k}\,,\qquad\tS_{k,N+1}\equiv\tS_{1k}\,.
\end{equation}
A straightforward calculation then shows that
\begin{eqnarray*}
  \fl
  T^\dagger\cH
  T&=\frac14\sum_{2\le i<j\le N}\bigg[\frac{1+S_{ij}}{\sin^2(\th_i-\th_j)}
     +\frac{1+\tS_{ij}}{\cos^2(\th_i-\th_j)}\bigg]
     +\frac14\sum_{j=2}^{N}\bigg[\frac{1+\tS_{1j}}{\sin^2(\th_j-\th_1)}
     +\frac{1+S_{1j}}{\cos^2(\th_j-\th_1)}\bigg]\\
  \fl
   &=  
     \cH-\sum_{j=2}^N\frac{\cos\Bigl(2(\th_j-\th_1)\Bigr)}{\sin^2\Bigl(2(\th_j-\th_1)\Bigr)}\,
     S_{1j}(1-S_1S_j)
\end{eqnarray*}
(cf.~Ref.~\cite{BFG20} for more details). Remarkably, a form of translation invariance can be
recovered by combining an ordinary translation with a spin flip in one of the chain's ends. More
precisely, let us define the \emph{twisted translation} operator $\cT$ by
\[
  \cT:=TS_1,
\]
whose action on the canonical spin basis~\eqref{basis} is given by
\[
  \cT\ket{s_1,\dots,s_N}=\ket{s_2,\dots,s_N,s_1'}\,.
\]
Note that $\cT$ is unitary, since both $T$ and $S_1$ are; indeed, $S_1$ is idempotent
($S_1^2=S_1$) and self-adjoint. It is straightforward to show that
\[
  \cT^\dagger S_{ij}\cT=S_{i+1,j+1},\qquad \cT^\dagger \tS_{ij}\cT=\tS_{i+1,j+1},
\]
provided that we set
\[
  S_{k,N+1}\equiv\tS_{1k},\qquad \tS_{k,N+1}\equiv S_{1k}.
\]
From these relations it readily follows that
\begin{eqnarray*}
  \cT^\dagger\cH \cT=\cH,
\end{eqnarray*}
so that the chain Hamiltonian~\eqref{Hchain} commutes with the elements of the group of twisted
translations generated by $\cT$. Note that the identity~$S_iT=TS_{i+1}$ implies that
$\cT^k=T^kS_k\cdots S_1$ for $k=1,\dots,N$. In particular,
\[
  \cT^N=S_N\cdots S_1\ne\id\,,\quad\text{but}\quad\cT^{2N}=\id.
\]
Thus the twisted translation group generated by the operator $\cT$ is a cyclic group of order
$2N$, i.e., \emph{twice} the order of the standard translation group generated by the ordinary
translation operator $T$.

\subsection{Boson-fermion duality}

We shall next analyze the relation between the Hamiltonians $\cH^{(m\vep_B|n\vep_F)}$ and
$\cH^{(n\vep_F|m\vep_B)}$ (which, by Remark~\ref{rem.vep}, coincides with
$\cH^{(n\vep_B|m\vep_F)}$), differing by the exchange of the bosonic and fermionic degrees of
freedom. To this end, we start by defining the operator $\chi^{(m|n)}:\cS^{(m|n)}\to\cS^{(n|m)}$
by
\[
  \chi^{(m|n)}\ket{s_1,\dots,s_N}=\ket{\hat s_1,\dots,\hat s_N},
\]
where
\[
  \hat s_i=\cases{s_i+n,& $m\in B$,\\ s_i-m,& $s_i\in F$.}
\]
In other words, the state $\ket{\hat s_1,\dots,\hat s_N}$ is obtained replacing the $k$-th bosonic
(resp.~fermionic) spin in $\ket{s_1,\dots,s_N}$ by the $k$-th fermionic (resp.~bosonic) spin.
Following Refs.~\cite{BBHS07,BFGR09}, we next define $\rho:\cS^{(m|n)}\to\cS^{(m|n)}$ by
\[
  \rho\ket{s_1,\dots,s_N}=(-1)^{\sum_k k\pi(s_k)}\ket{s_1,\dots,s_N},
\]
where
\[
  \pi(s_i)=\cases{
    0,& $s_i\in B$,\\
    1,& $s_i\in F$, }
\]
and consider the operator
\[
  U:=\chi^{(m|n)}\rho:\cS^{(m|n)}\to\cS^{(n|m)}.
\]
The operator $U$ is clearly unitary, since it maps one orthonormal basis into another one.
Moreover, since obviously $\rho^{-1}=\rho$ and $\big(\chi^{(m|n)}\big)^{-1}=\chi^{(n|m)}$ it
follows that $U^{-1}=\rho\chi^{(n|m)}$. We then have
\begin{eqnarray*}
  \fl
  U^{-1}S_{ij}^{(n|m)}U\ket{s_1,\dots,s_N}
  &=(-1)^{\sum_k k\pi(s_k)}\rho\chi^{(n|m)}S_{ij}^{(n|m)}
    \ket{\hat s_1,\dots,\hat s_N}\\
  \fl
  &=(-1)^{\sum_k k\pi(s_k)}(-1)^{\nu(\hat s_i,\dots,\hat s_j)}\rho\chi^{(n|m)}
    \ket{\hat s_1,\dots,\hat s_j,\dots,\hat s_i,\dots,\hat s_N}\\
  \fl
  &=\vep(s_i,\dots,s_j)\ket{s_1,\dots,s_j,\dots,s_i,\dots, s_N}
\end{eqnarray*}
with
\begin{eqnarray*}
  \fl
  \vep(s_i,\dots,s_j)
  &=(-1)^{\sum_k k\pi(s_k)}(-1)^{\sum_{k\ne i,j} k\pi(s_k)}(-1)^{i\pi(s_j)+j\pi(s_i)}(-1)^{\nu(\hat
    s_i,\dots,\hat s_j)}\\
  &=(-1)^{(j-i)(\pi(s_i)-\pi(s_j))} (-1)^{\nu(\hat
    s_i,\dots,\hat s_j)}.
\end{eqnarray*}
Clearly, when $\pi(s_i)=\pi(s_j)$ we have
\[
  \vep(s_i,\dots,s_j)=(-1)^{\nu(\hat s_i,\dots,\hat s_j)}=(-1)^{\pi(\hat s_i)}=-(-1)^{\pi(s_i)}
  =-(-1)^{\nu(s_i,\dots,s_j)}.
\]
On the other hand, when $\pi(s_i)\ne\pi(s_j)$ the identity
\[
  \nu(s_i,\dots,s_j)+\nu(\hat s_i,\dots,\hat s_j)=j-i-1
\]
implies that also in this case $\vep(s_i,\dots,s_j)=-(-1)^{\nu(s_i,\dots,s_j)}$. Hence
\[
  \fl
  U^{-1}S_{ij}^{(n|m)}U\ket{s_1,\dots,s_N}=-(-1)^{\nu(s_i,\dots,s_j)}\ket{s_1,\dots,s_j,\dots,s_i,\dots,
    s_N} =-S_{ij}^{(m|n)}\ket{s_1,\dots,s_N},
\]
and therefore
\begin{equation}\label{USij}
  U^{-1}S_{ij}^{(n|m)}U=-S_{ij}^{(m|n)}.
\end{equation}
Similarly,
\begin{eqnarray*}
  \fl
  U^{-1}S_{i}^{(n|m)}U\ket{s_1,\dots,s_N}
  &=(-1)^{\sum_k k\pi(s_k)}\rho\chi^{(n|m)}S_{i}^{(n|m)}
    \ket{\hat s_1,\dots,\hat s_N}\\
  \fl
  &=(-1)^{\sum_k k\pi(s_k)}\si(\hat s_i)\rho\chi^{(n|m)}
    \ket{\hat s_1,\dots,(\hat s_i)',\dots,\hat s_N}\\
  \fl
  &=\si(\hat s_i)\ket{s_1,\dots,s_i',\dots, s_N}=
    \si(s_i)\ket{s_1,\dots,s_i',\dots, s_N}\\
  &=S_i^{(m|n)}\ket{s_1,\dots,s_i,\dots, s_N},
\end{eqnarray*}
and thus
\begin{equation}
  \label{USi}
  U^{-1}S_{i}^{(n|m)}U=S_{i}^{(m|n)}.
\end{equation}
Combining Eqs.~\eqref{USij} and \eqref{USi} we obtain
\[
  U^{-1}\tS_{ij}^{(n|m)}U=-\tS_{ij}^{(m|n)},
\]
whence
\begin{equation}
  \fl
  U^{-1}\cH^{(n|m)}U
  =\frac14\sum_{i<j}\left(\frac{1+S_{ij}^{(m|n)}}{\sin^2(\th_i-\th_j)}
    +\frac{1+\tS_{ij}^{(m|n)}}{\cos^2(\th_i-\th_j)}\right)
  =\cE_0-\cH^{(m|n)}\,,
\end{equation}
where
\begin{equation}\label{cE0}
  \fl \cE_0=\frac12\sum_{i<j}\Big[\sin^{-2}(\th_i-\th_j)+\cos^{-2}(\th_i-\th_j)\Big]
  =2\sum_{i<j}\sin^{-2}\biggl(\frac{\pi(j-i)}N\biggr)=\frac13N(N^2-1)
\end{equation}
(see~Ref.~\cite{FG05} for the evaluation of the sum). In other words, the partition functions of
$\cH^{(n|m)}$ and $\cH^{(m|n)}$ are related by
\begin{equation}\label{Znmmn}
  \cZ^{(n|m)}(q)=q^{\cE_0}\cZ^{(m|n)}(q^{-1}).
\end{equation}
In particular, for $m=n$ we have
\begin{equation}\label{Zmm}
  \cZ^{(m|m)}(q)=q^{\cE_0}\cZ^{(m|m)}(q^{-1}),
\end{equation}
implying that the spectrum of $\cH^{(m|m)}$ is symmetric about $\cE_0/2$. Note that both
equations~\eqref{Znmmn} and~\eqref{Zmm} can be explicitly checked in the cases
$(m,n)=(0,2),(2,0),(2,2)$ studied in Section~\ref{sec.ee}.

\section{Thermodynamics}\label{sec.thermo}

With the help of the partition function $\cZ_N(q)$ of the $\su(m|n)$ chain~\eqref{Hchain} with $N$
spins computed in Section~\ref{sec.PF} one can in principle obtain the thermodynamic free energy
per particle
\begin{equation}\label{fdef}
  f(T)=-T\lim_{N\to\infty}N^{-1}\log\cZ_N(q),\qquad q=\e^{-1/T}\equiv\e^{-\be},
\end{equation}
from which all the other thermodynamic functions (internal energy, specific heat at constant
volume, entropy per particle, etc.) can be derived through the usual formulas
\[
  u=\frac{\pd}{\pd\be}(\be f),\qquad c_V=-\be^2\frac{\pd u}{\pd\be},\qquad s=\be^2\frac{\pd
    f}{\pd\be}=\be(u-f).
\]
In fact, for the previous formulas to make sense we must first normalize the
Hamiltonian~\eqref{Hchain} so that the average energy per spin tends to a finite constant in the
thermodynamic limit $N\to\infty$. The average energy $\langle\cH\rangle=(m+n)^{-N}\tr\cH$ of the
Hamiltonian~\eqref{Hchain} is easily computed by taking into account the identities
\[
  \tr S_{ij}=\tr(\tS_{ij})=(m+n)^{N-2}(m-n),
\]
whence
\[
  \fl
  \langle\cH\rangle=\left(1+\frac{n-m}{(m+n)^2}\right)\sum_{i<j}\sin^{-2}\bigl(\pi(i-j)/N\bigr)
  =\frac16 N(N^2-1)\left(1+\frac{n-m}{(m+n)^2}\right)
\]
(see Ref.~\cite{FG05} for the evaluation of the sum). Note that this is twice the value of the
average energy of the $\su(m|n)$ HS chain~\cite{BB06}. Thus in this section we shall take
\begin{equation}\label{Hnorm}
   \cH=\frac{J}{4N^2}\sum_{1\le i<j\le N}\left(\frac{1-S_{ij}}{\sin^2(\th_i-\th_j)}
    +\frac{1-\tS_{ij}}{\cos^2(\th_i-\th_j)}\right),
\end{equation}
whose average energy per spin $\langle\cH\rangle/N$ tends to a constant in the thermodynamic
limit. Note that the constant $J$, which sets the energy scale, could be of either sign.

\begin{figure}[t]
  \includegraphics[width=.48\linewidth]{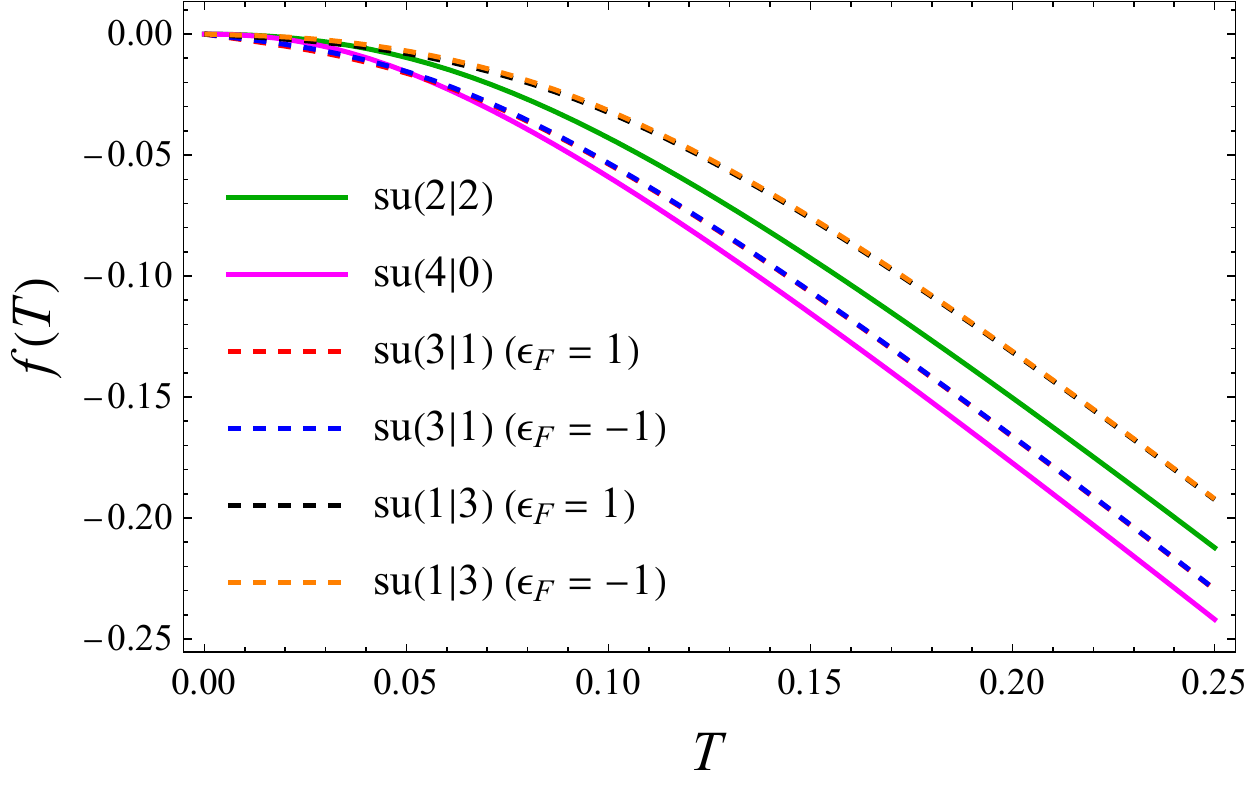}\hfill\includegraphics[width=.48\linewidth]{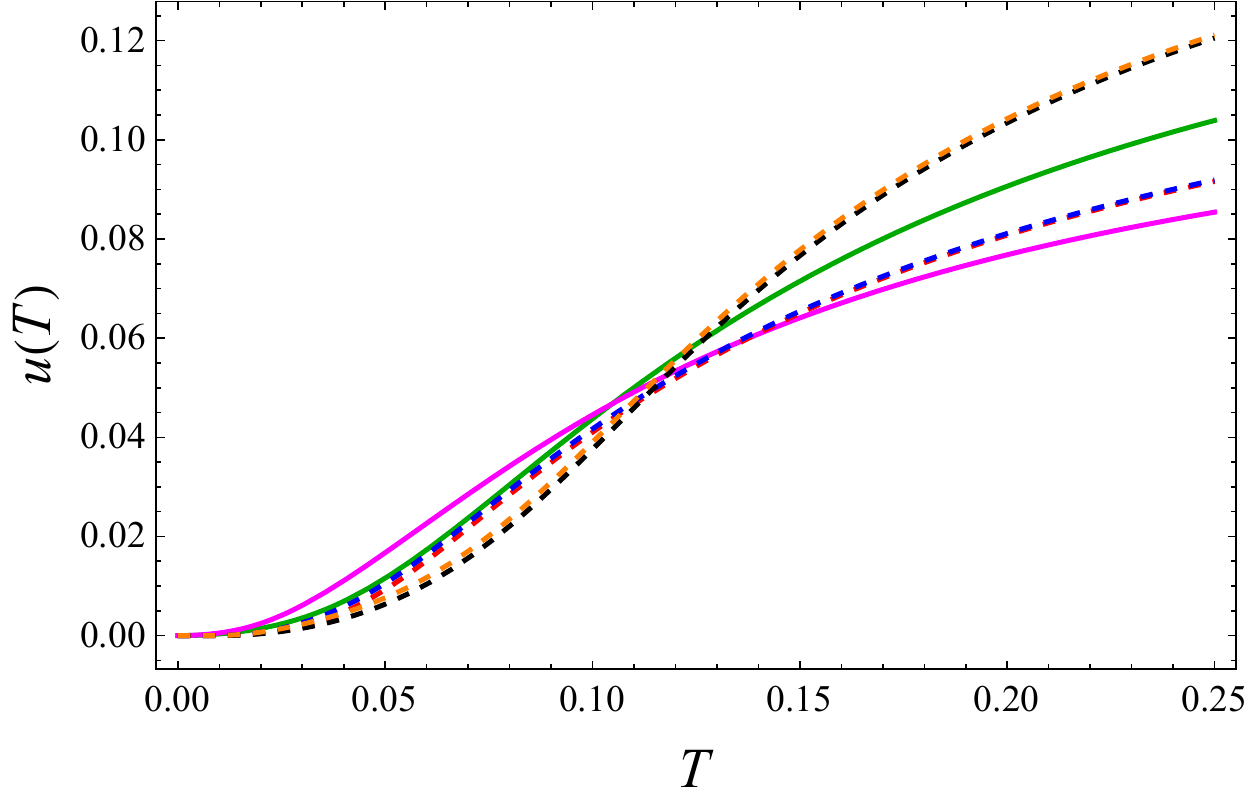}\\
  \includegraphics[width=.48\linewidth]{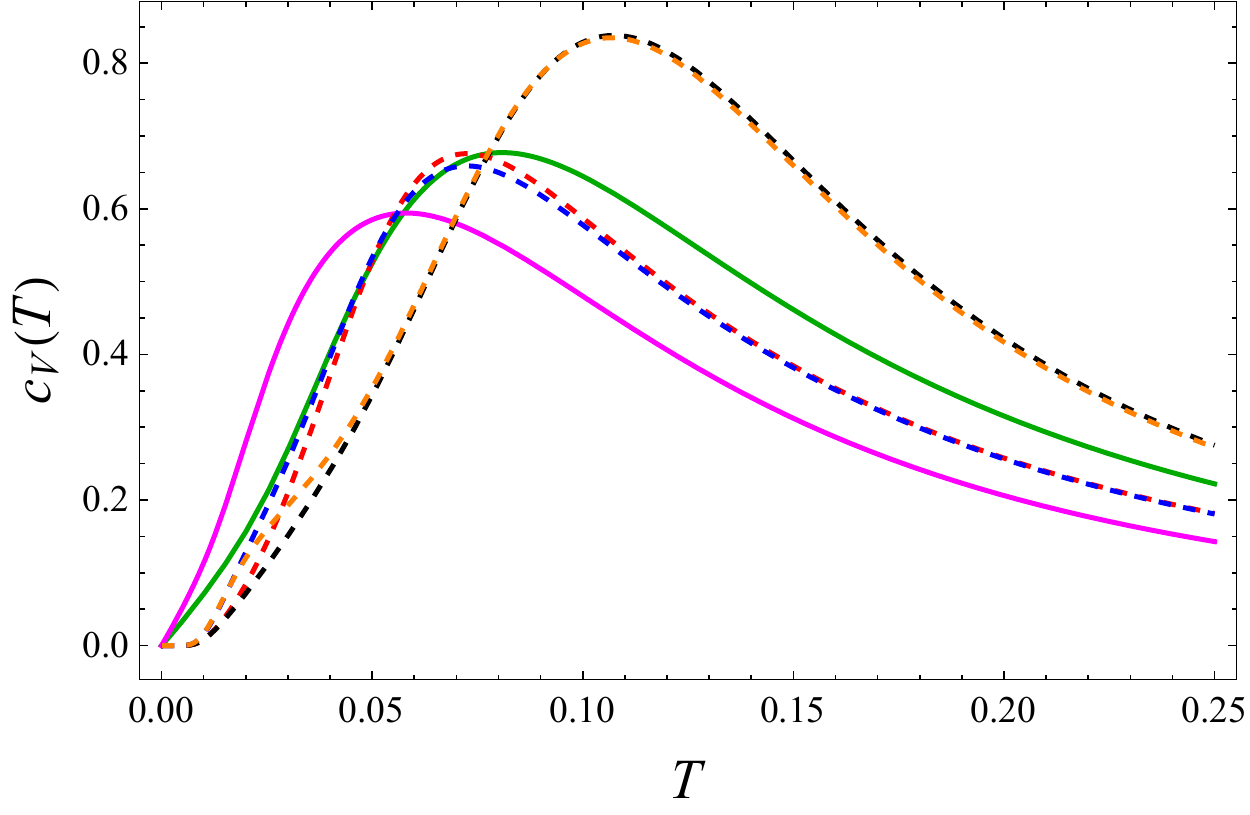}\hfill\includegraphics[width=.48\linewidth]{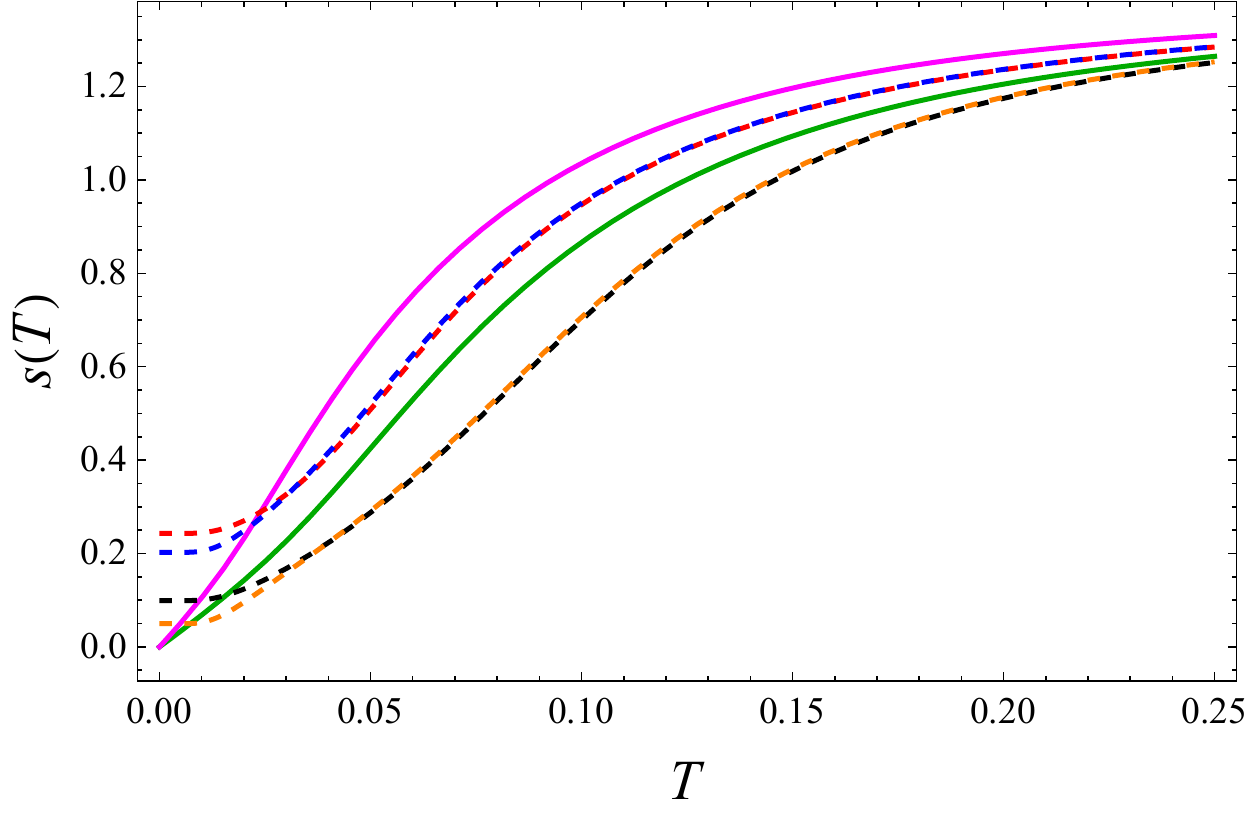}
  \caption{Thermodynamic functions of the $\su(m|n)$ chains~\eqref{Hnorm} with $m+n=4$ (the
    $\su(0|4)$ chain has been omitted since, as we shall show below, it is thermodynamically
    equivalent to the $\su(4|0)$ chain). The color code is as indicated in the plot of $f$ (top
    left), and in all plots the temperature is measured in units of $J$. The thermodynamic
    functions of the $\su(3|1)$ and $\su(1|3)$ chains, for whose $N\to\infty$ limit there is no
    known closed-form expression, have been computed using the exact partition function for $N=14$
    spins.}
  \label{fig.sumn4}
\end{figure}

In practice, except in the case of even $m$
and $n$ that we shall discuss below, the complexity of the expressions for $\cZ_N$ makes it
unfeasible to compute the thermodynamic free energy in closed form. However, even in these cases
it is possible to obtain an approximation of the thermodynamic free energy through the formula
\begin{equation}\label{fTN}
  f(T)\simeq f_N(T)=-\frac TN\,\log\cZ_N(q)
\end{equation}
with a sufficiently large $N$. For instance, in Fig.~\ref{fig.sumn4} we present the plots of the
thermodynamic functions of the $\su(3|1)$ and $\su(1|3)$ chains computed from Eq.~\eqref{fTN} with
$N=14$, compared to their analogues for the $\su(2|2)$ and $\su(4|0)$ chains in the thermodynamic
limit (cf.~Eqs.~\eqref{su22} and~\eqref{um0}--\eqref{sm0}). In particular, this and similar plots
suggest that in the thermodynamic limit the thermodynamics of the $\su(3|1)$ and $\su(1|3)$ chains
are independent of $\vep_F$.
\begin{remark}
  From the duality relation~\eqref{Znmmn} it is immediate to obtain a relation between the
  thermodynamic functions of the $\su(m|n)$ and $\su(n|m)$ chains with opposite values of the
  parameter $J$. Indeed, we can rewrite Eq.~\eqref{Zmm} more explicitly as
\[
  \cZ_N^{(m|n)}(T;-J)=\e^{J\be\cE_0/N^2}\cZ_N^{(n|m)}(T;J),
\]
where the factor of $J/N^2$ in the exponential is due to the new normalization~\eqref{Hnorm} of
the chain's Hamiltonian. From Eqs.~\eqref{cE0} and~\eqref{fdef} we then obtain
\begin{equation}\label{dualrel}
  \fl
  f^{(m|n)}(T;-J)=f^{(n|m)}(T;J)
  -\frac{J}3\lim_{N\to\infty}\left(1-\frac1{N^2}\right)=f^{(n|m)}(T;J)-\frac{J}3.
\end{equation}
For this reason, in what follows we shall restrict ourselves without loss of generality to the
case $J>0$. In particular, for $m=n$ the previous formula becomes
\begin{equation}\label{fmm}
   f^{(m|m)}(T;-J)=f^{(m|m)}(T;J)-\frac{J}3,
 \end{equation}
 from which it easily follows that
 \begin{eqnarray}
   u^{(m|m)}(T;-J)&=u^{(m|m)}(T;J)-\frac{J}3,\label{umm}\\
   c_V^{(m|m)}(T;-J)&=c_V^{(m|m)}(T;J),\label{cVmm}\\
   s^{(m|m)}(T;-J)&=s^{(m|m)}(T;J).\label{smm}
 \end{eqnarray}
 Hence in this case the $J>0$ and $J<0$ cases are actually equivalent.\qed
\end{remark}

As mentioned above, when $m$ and $n$ are both even the factorization~\eqref{fact} of the partition
function makes it possible to express the thermodynamic free energy per spin $f^{(m|n)}$ of the
$\su(m|n)$ chain~\eqref{Hnorm} in terms of the free energies per spin of the $\su(1|1)$ and
$\su(\frac m2|\frac n2)$ Haldane--Shastry chains as
\begin{equation}\label{fmnfHS}
  f^{(m|n)}(T)=f_\HS^{(1|1)}(T)+f_\HS^{(\frac m2|\frac n2)}(T).
\end{equation}
In fact, a closed form expression for the free energy per spin $f_\HS^{(p|q)}(T)$ of the
$\su(p|q)$ HS chain was developed in Ref.~\cite{FGLR18}, namely
\[
  f_\HS^{(p|q)}(T)=-T\int_0^1\log\la^{(p|q)}(x)\,\diff x,
\]
where $\la^{(p|q)}(x)$ is the Perron--Frobenius (i.e., largest in modulus) eigenvalue of the
transfer matrix $A^{(p|q)}(x)$ of order $m+n$ introduced in Ref.~\cite{FGLR18}. More precisely, the
matrix elements of $A^{(p|q)}(x)$ are defined by
\[
A^{(p|q)}_{\al\ga}(x)=\e^{-\be J\rho(x)\de(\al,\ga)},\qquad 1\le\al,\ga\le m+n,
\]
where
\[
  \rho(x)=x(1-x)
\]
is the continuous version of the dispersion relation~$\cE(i)/N^2$ (with $i/N\to x$) and
\[
  \de(\al,\ga)= \cases{
    0,& $\al<\ga$\en or\en$\al=\ga\in B$,\\
    1,& $\al>\ga$\en or\en$\al=\ga\in F$. }
\]
For instance, for $p=q=1$ we have
\[
  A^{(1|1)}(x)=
  \left(
  \begin{array}{cc}
    1& 1\\
    \e^{-\be J\rho(x)}&\e^{-\be J\rho(x)}
  \end{array}
  \right)
\]
so that
\[
  \la^{(1|1)}(x)=1+\e^{-\be J\rho(x)}.
\]
On the other hand, the Perron--Frobenius eigenvalue for the case $q=0$ was found in
Ref.~\cite{FG22pre} to be
\begin{equation*}
  \la^{(p|0)}(x)=\frac{1-\e^{-\be J\rho(x)}}{1-\e^{-\be J\rho(x)/p}}=\e^{(1-p)\be J\rho(x)/(2p)}
                  \frac{\sinh\Big(\be J\rho(x)/2\Big)}{\sinh\Big(\be J\rho(x)/(2p)\Big)}.
\end{equation*}
Thus when $m$ is even the free energy of the chain~\eqref{Hnorm} in the non-supersymmetric case
$n=0$ is explicitly given by
\begin{equation}\label{fm0}
  \fl f^{(m|0)}(T)=\frac{J}6\left(1-\frac1{m}\right)-T\int_0^1\log \left(\frac{\sinh\Big(\be
      J\rho(x)\Big)}{\sinh\Big(\be J\rho(x)/m\Big)}\right)\diff x.
\end{equation}
Note also that, since the integral in Eq.~\eqref{fm0} does not depend on the sign of $J$, for even
$m$ we have
\begin{eqnarray*}
  f^{(m|0)}(T;-J)&=-\frac{J}3\left(1-\frac1{m}\right)+f^{(m|0)}(T;J),\\
  u^{(m|0)}(T;-J)&=-\frac{J}3\left(1-\frac1{m}\right)+u^{(m|0)}(T;J),\\
  c_V^{(m|0)}(T;-J)&=c_V^{(m|0)}(T;J),\qquad s^{(m|0)}(T;-J)=s^{(m|0)}(T;J).
\end{eqnarray*}
In particular combining the previous equation for $f^{(m|0)}(T;-J)$ with the duality
relation~\eqref{dualrel} we deduce that
\[
  f^{(0|m)}(T;J)=f^{(m|0)}(T;-J)+\frac{J}3=f^{(m|0)}(T;J)+\frac{J}{3m}.
\]
Hence the $\su(m|0)$ and $\su(0|m)$ cases are thermodynamically equivalent.

Interestingly, the free energy~\eqref{fm0} of the $\su(m|0)$ chain with even $m$ coincides with
the free energy of the ordinary (bosonic) $\su(m|0)$ HS chain with $J$ replaced by $2J$. In other
words, both models are equivalent in the thermodynamic limit up to a suitable rescaling of the
coupling constant $J$. By the remark preceding Eq.~\eqref{Hgen} in Section~\ref{sec.model}, we
thus obtain in this case the same thermodynamics as if we replaced the spin flip operators $S_i$
by plus or minus the identity. For the sake of completeness, we list below the explicit formulas
for the internal energy, specific heat at constant volume and entropy per particle of the
$\su(m|0)$ chain with even $m$ (the corresponding formulas for the $\su(0|n)$ chain with even $n$
are similar and shall therefore be omitted):
\begin{eqnarray}
  \fl
  u^{(m|0)}(T)&=\frac{J}6\left(1-\frac1{m}\right)-J\int_0^1\rho(x)\left[\coth(\be J\rho(x))
                -\frac1m\coth(\be J\rho(x)/m)\right]\diff x,\label{um0}\\\fl
  c_V^{(m|0)}(T)&=\be^2J^2\int_0^1\rho^2(x)\left[\frac1{m^2}\csch^2(\be J\rho(x)/m)
                  -\csch^2(\be J\rho(x))\right]\diff x,\label{cvm0}\\\fl
  s^{(m|0)}(T)&=\int_0^1\left[\log \left(\frac{\sinh\Big(\be
                J\rho(x)\Big)}{\sinh\Big(\be J\rho(x)/m\Big)}\right)\right.\nonumber\\\fl
                &\kern 5em-\be J\rho(x)\left(\coth(\be J\rho(x))
                  -\frac1m\coth(\be J\rho(x)/m)\right)
                  \left.\vphantom{\log \left(\frac{\sinh\Big(\be
                  J\rho(x)\Big)}{2\sinh\Big(\be J\rho(x)/m\Big)}\right)}
                  \right]\diff x.\label{sm0}
\end{eqnarray}
\begin{figure}[t]
  \includegraphics[width=.48\linewidth]{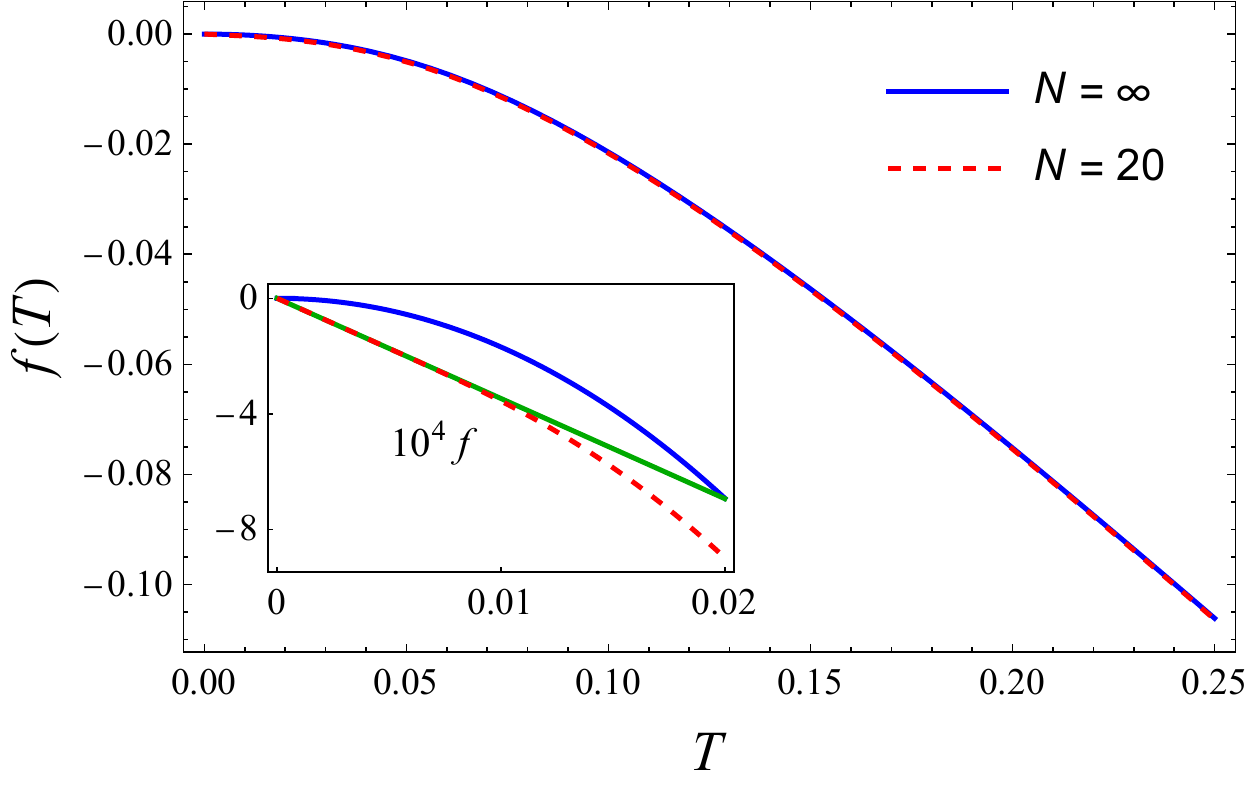}\hfill
  \includegraphics[width=.48\linewidth]{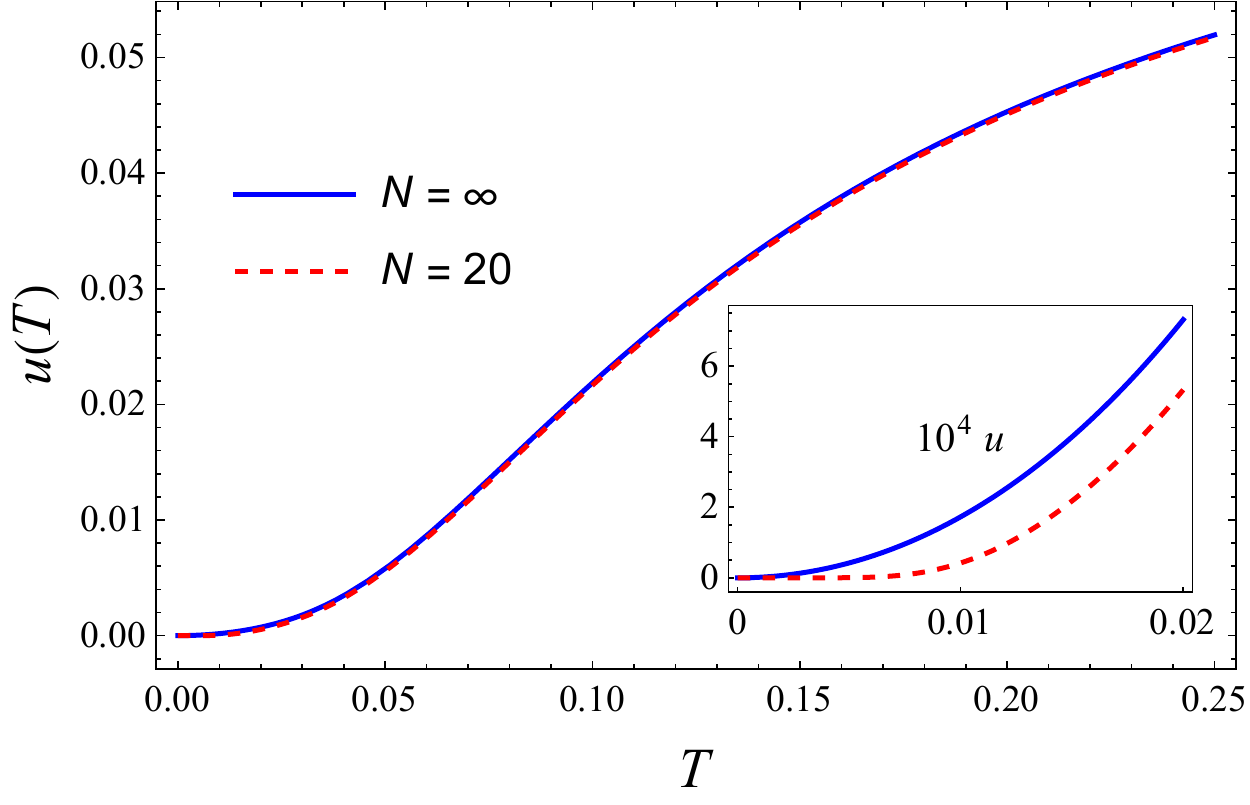}\\
  \includegraphics[width=.48\linewidth]{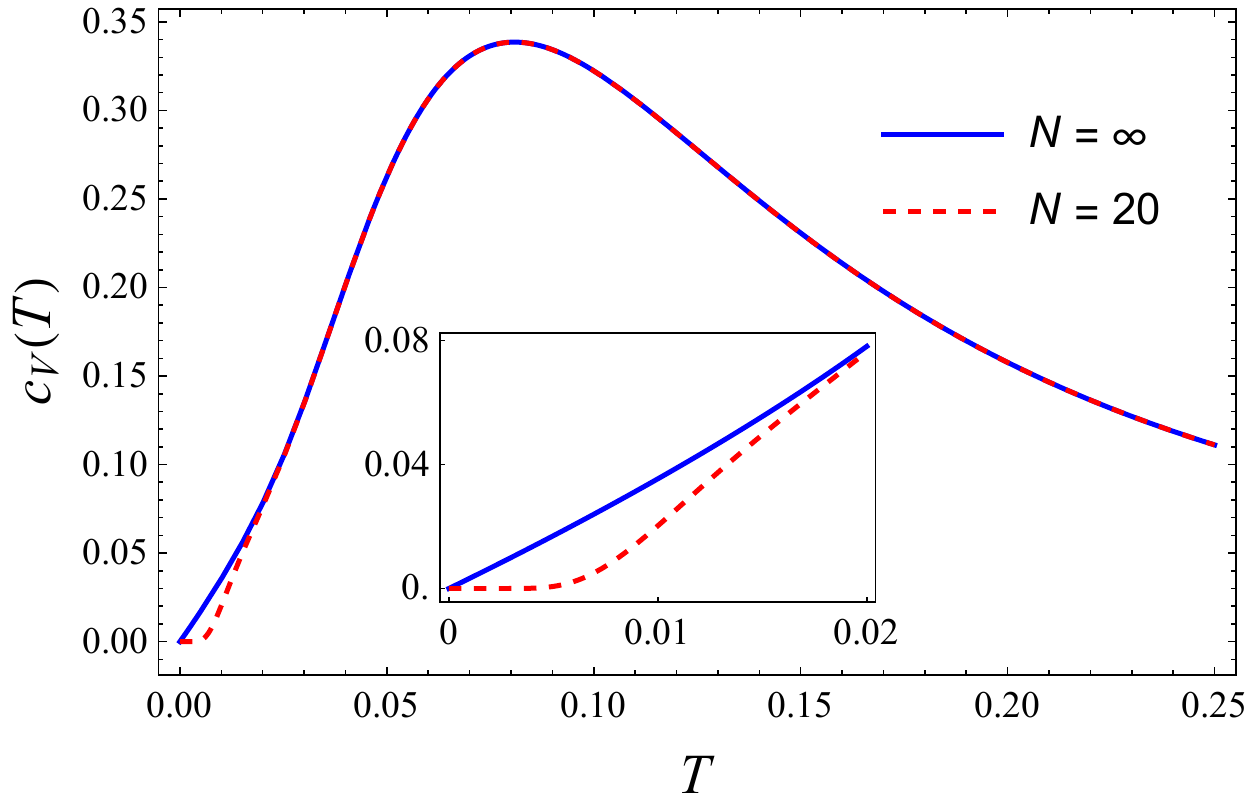}\hfill
  \includegraphics[width=.48\linewidth]{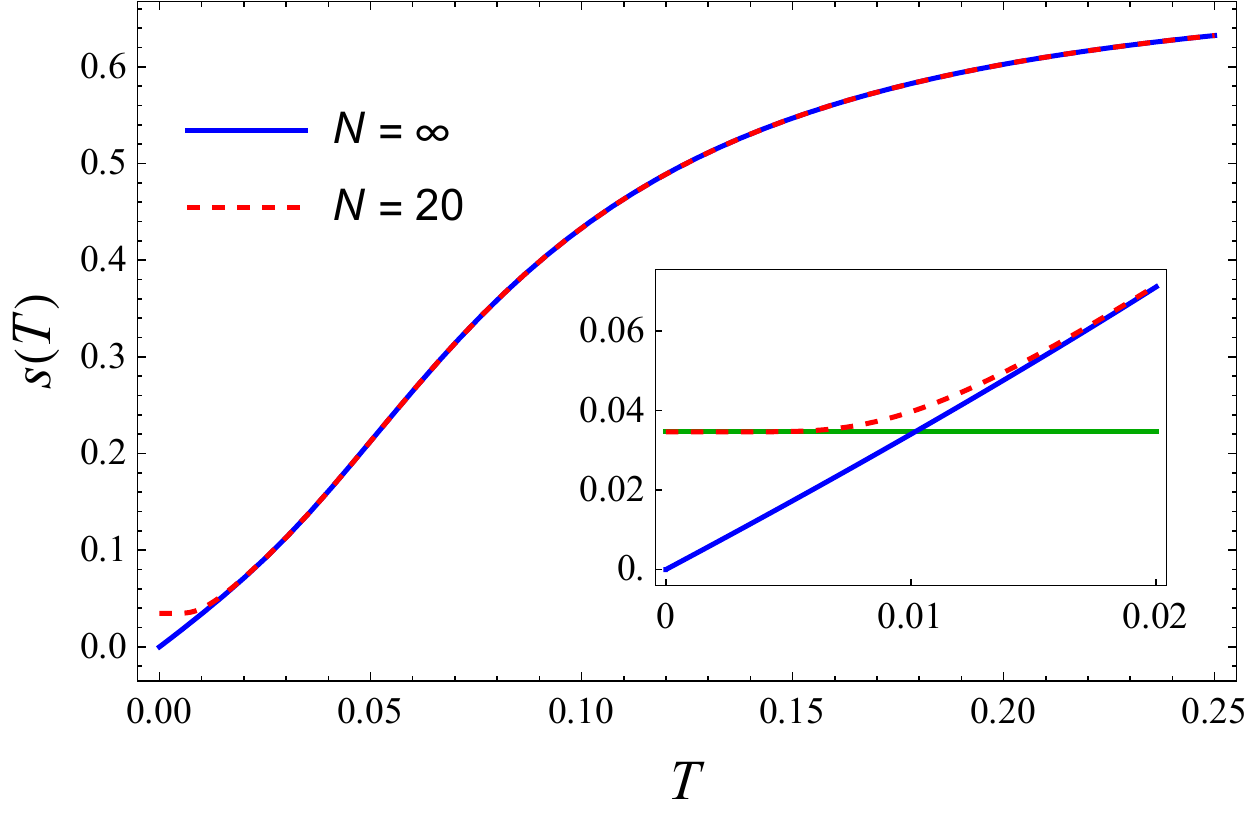}
    \caption{Thermodynamic functions of the $\su(2|0)$ chain~\eqref{Hnorm} with $N=20$ spins (dashed
    red line) and its thermodynamic limit (continuous blue line). In all plots the temperature is
    measured in units of $J$, and the insets show the low temperature behavior of each
    thermodynamic function. The significance of the continuous green lines in the insets of the
    plots of the free energy (top left) and the entropy (bottom left) is explained in the body of
    the article.}
  \label{fig.su20}
\end{figure}

In Fig.~\ref{fig.su20} we plot the thermodynamic functions of the $\su(2|0)$ chain for $N=20$
spins, compared to their thermodynamic limits given by Eqs.~\eqref{fm0}-\eqref{sm0}. We see that
the agreement between both plots is excellent even for this relatively low number of spins, except
at very low temperatures (which in our units corresponds to $T\lesssim 2\cdot 10^{-2}$). This low
temperature discrepancy is, in fact, an inevitable finite size effect. Indeed, at very low
temperatures we have
\[
  \cZ_N(T)\simeq \dGS\e^{-\be\EGS}+d_1\e^{-\be E_1},
\]
where $\EGS$ and $E_1=\EGS+\De E$ denote respectively the energies of the ground state and the
first excited state, and $\dGS$ and $d_1$ are their degeneracies. Thus for finite $N$ the chain's
thermodynamic functions behave at low temperatures as
\begin{eqnarray*}
  f_N(T)&\simeq \frac{\EGS}N-\frac{\log\dGS}{N}\,T-\frac{d_1}{\dGS}\frac{T}N\,\e^{-\be\De E},\\
  u_N(T)&\simeq \frac{\EGS}N+\frac{\De E}N\frac{d_1}{\dGS}\,\e^{-\be\De E},\\
  c_{V,N}(T)&\simeq \frac{d_1}{\dGS}\frac{(\De E)^2}{NT^2}\,\e^{-\be\De E},\\
  s_N(T)&\simeq \frac{\log\dGS}{N}+\frac{d_1}{\dGS}\frac{\De E}{NT}\,\e^{-\be\De E}.
\end{eqnarray*}
In our case ($m=2$, $n=0$) we have $\EGS=0$ and $\dGS=2$, so that at low temperatures the free
energy for finite $N$ is approximately a linear function of the temperature with slope equal to
$-(\log2)/N$. This is corroborated by the inset in the upper left corner of Fig.~\ref{fig.su20},
where the line $f=-(\log2)T/20$ has been plotted in green. It is also clear from the previous
equations that the remaining thermodynamic functions differ from their zero temperature values by
exponentially small terms of the form $T^{-k}\e^{-\be\De E}$ for suitable $k\in\{0,1,2\}$; this
behavior is again apparent from the corresponding insets in Fig.~\ref{fig.su20} (note that
$\De E=19$ and $d_1=4$ for $N=20$). In particular, the zero temperature value of the entropy for a
finite number of spins $N$ is
\[
  s_N(0)=\frac{\log2}{N}\ne0\,;
\]
this can be seen from the bottom left inset of Fig.~\ref{fig.su20}, where we have plotted in green
the horizontal line $s=(\log 2)/20\simeq0.0346574$.

Apart from the previous non-supersymmetric cases, it is also possible to evaluate the
thermodynamic functions of the chain~\eqref{Hnorm} in a few genuinely supersymmetric cases with
$m$ and $n$ even using the results of Ref.~\cite{FGLR18}. To begin with, for $m=n=2$ we have
\begin{equation}\label{su22}
  \fl
  f^{(2|2)}(T;J)=2f^{(1|1)}_{\HS}(T;J)=-2T\int_0^1\log\big(1+\e^{-\be J\rho(x)}\big)\diff x=
  f^{(2|2)}_{\HS}(T;2J),
\end{equation}
where we have used the explicit expression for $f^{(2|2)}_{\HS}$ in Ref.~\cite{FGLR18}. Thus the
thermodynamics of the $\su(2|2)$ chain is equivalent to that of its HS counterpart with $J$
replaced by $2J$, obtained by setting $S_i=1$ in Eq.~\eqref{Hnorm}. Likewise, for $m=4$, $n=2$,
using the expression of $f^{(2|1)}_\HS$ in Ref.~\cite{FGLR18} we obtain
\begin{eqnarray*}
  f^{(4|2)}(T)&=f^{(1|1)}_\HS(T)+f^{(2|1)}_\HS(T)\\
              &=
                -T\int_0^1\log\big(1+\e^{-\be J\rho(x)}\big)\diff x\\
              &\hphantom{=-T\int}
                -T\int_0^1\log\bigg[1+\frac12\e^{-\be J\rho(x)}+\frac12\e^{-\be J\rho(x)}
                \sqrt{1+8\e^{\be J\rho(x)}}\,\bigg]\diff x.
\end{eqnarray*}
Although it is straightforward to obtain the energy, specific heat at constant volume and entropy
from the previous formula, the resulting expressions are somewhat lengthy and shall therefore be
omitted. The free energy of the $\su(2|4)$ case is easily obtained from the latter equation for
$f^{(4|2)}$ using the duality relation~\eqref{dualrel}:
\begin{eqnarray}
   f^{(2|4)}(T)&=
                -T\int_0^1\log\big(1+\e^{-\be J\rho(x)}\big)\diff x\nonumber\\
              &\hphantom{=-T\int}
                -T\int_0^1\log\bigg[\frac12+\e^{-\be J\rho(x)}+\frac12
                \sqrt{1+8\e^{-\be J\rho(x)}}\,\bigg]\diff x.
                \label{fsu24}
\end{eqnarray}
Finally, the thermodynamic free energy per particle for the case $m=n=4$ can also be
computed in closed form using the results of Ref.~\cite{FGLR18}, namely
\begin{eqnarray*}
  f^{(4|4)}(T)&=f^{(1|1)}_\HS(T)+f^{(2|2)}_\HS(T)\\
  &=
  -T\int_0^1\log\big(1+\e^{-\be J\rho(x)}\big)\diff x
    -2T\int_0^1\log\big(1+\e^{-\be J\rho(x)/2}\big)\diff x\\
              &=\frac{J}6-T\int_0^1\log\bigg(8\cosh\left(\tfrac{\be J\rho(x)}2\right)
               \cosh^2\left(\tfrac{\be J\rho(x)}4\right) \bigg)\diff x
  \\
    &\equiv f^{(1|1)}_\HS(T)+f^{(1|1)}_\HS(2T).
\end{eqnarray*}
\begin{figure}[t]
  \includegraphics[width=.48\linewidth]{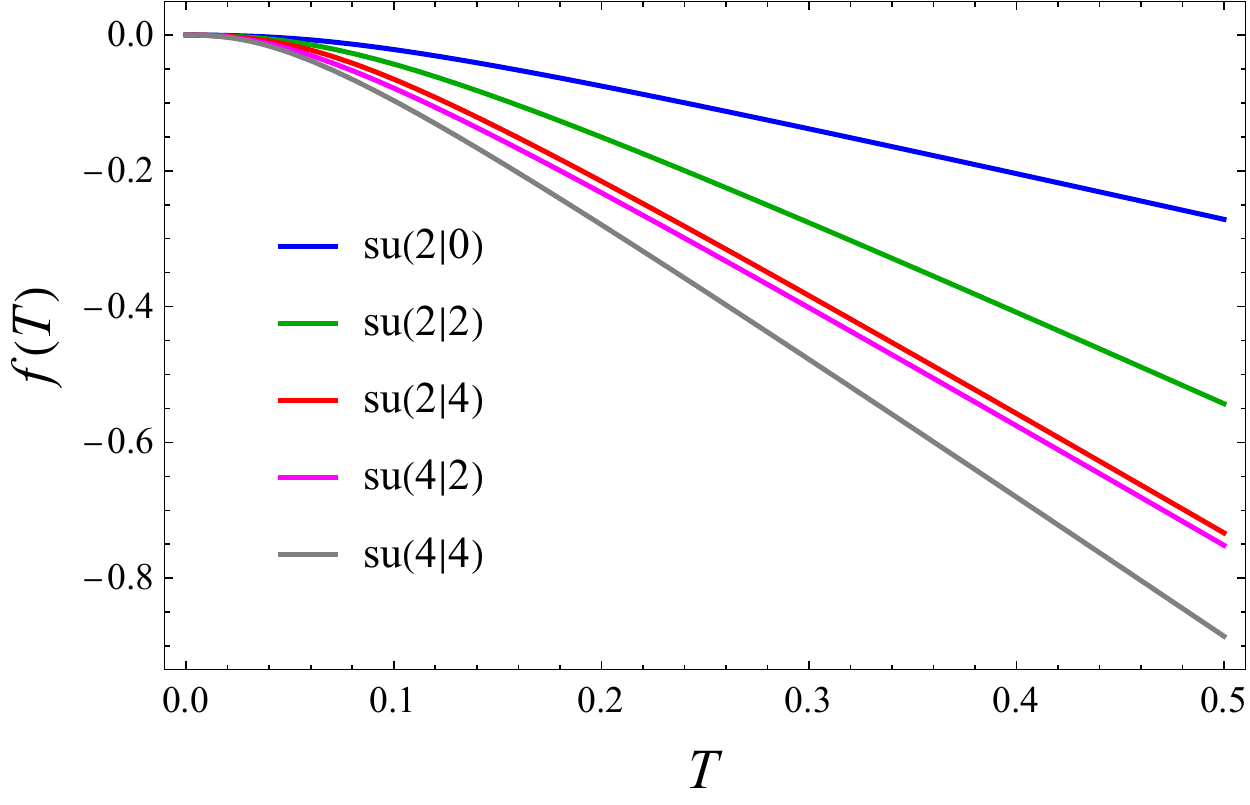}\hfill
  \includegraphics[width=.48\linewidth]{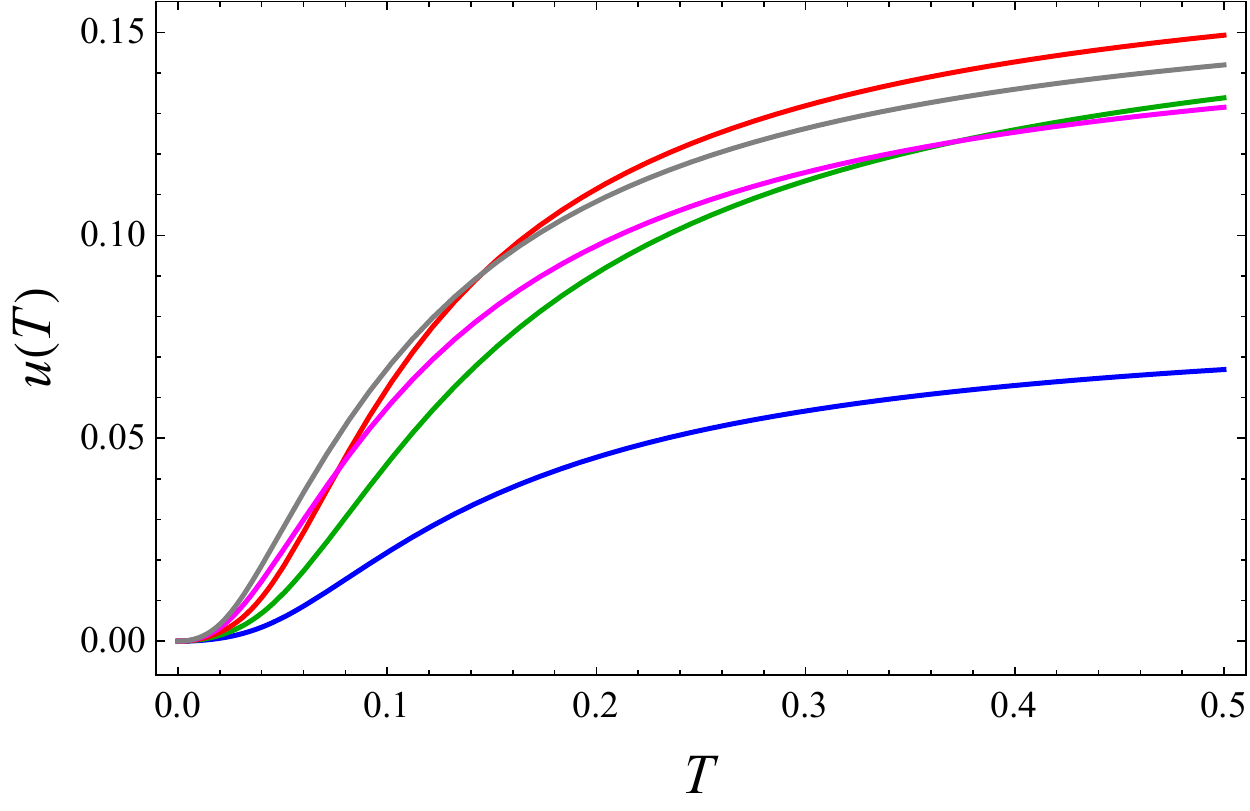}\\
    \includegraphics[width=.48\linewidth]{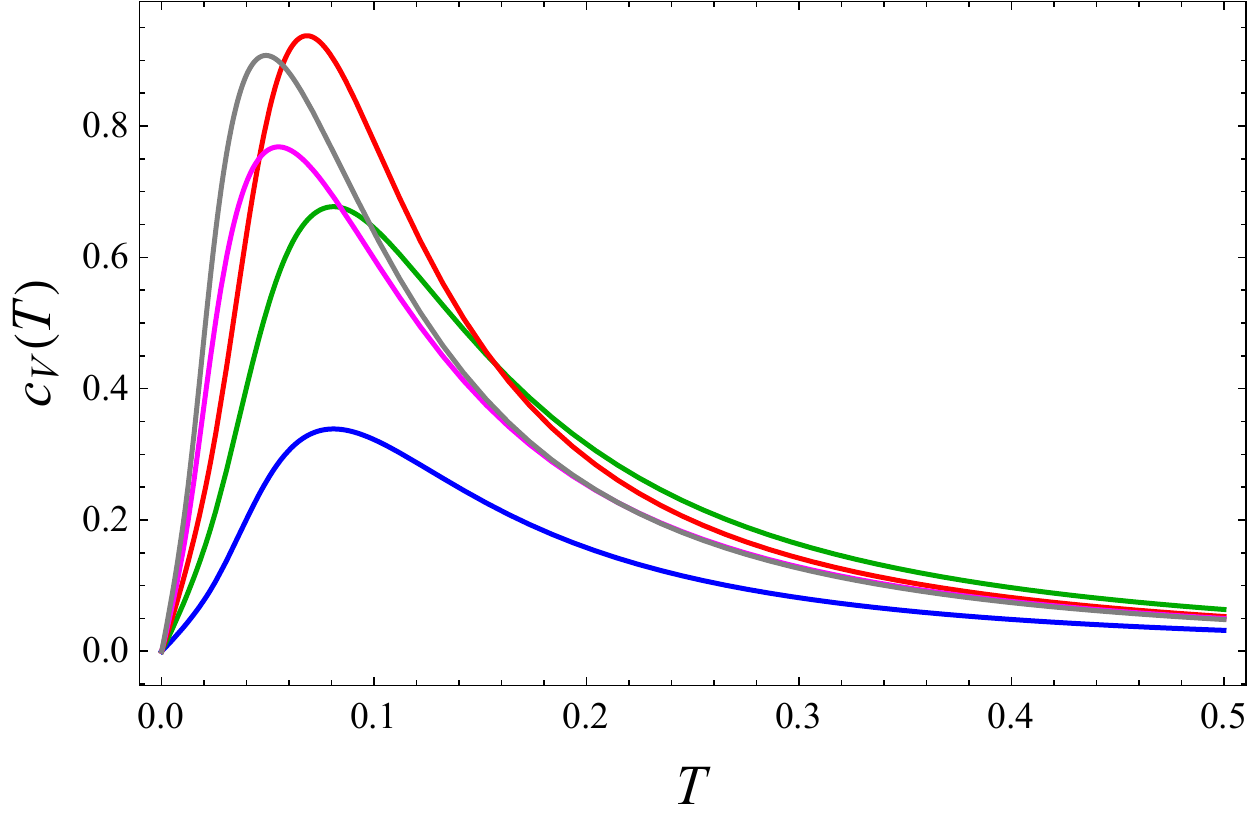}\hfill
  \includegraphics[width=.48\linewidth]{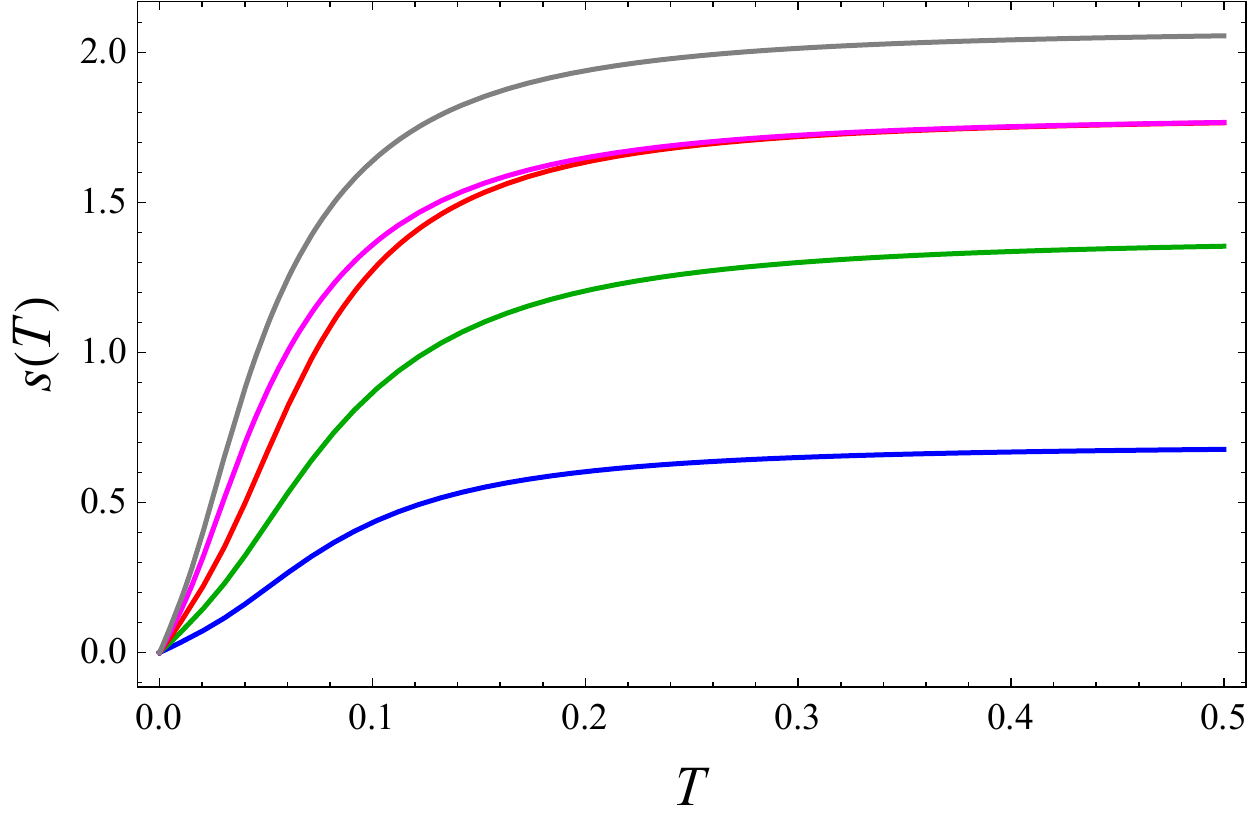}
  \caption{Thermodynamic functions of the $\su(2|2)$, $\su(2|4)$, $\su(4|2)$ and $\su(4|4)$
    chains~\eqref{Hnorm}, whose thermodynamic functions have closed-form expressions in the
    thermodynamic limit (color code in top left figure). In all plots the temperature is measured
    in units of $J$. The thermodynamic functions of the non-supersymmetric $\su(2|0)$ chain are
    also shown for comparison purposes.}
  \label{fig.sumnexact}
\end{figure}%
In this case the expressions for the remaining thermodynamic functions are relatively simple, to
wit
\begin{eqnarray*}
  \fl
  u^{(4|4)}(T)&=J\int_0^1\left(\frac1{1+\e^{\be J\rho(x)}}+\frac1{1+\e^{\be
                J\rho(x)/2}}\right)\rho(x)\diff x,\\
  \fl
              &=\frac{J}6-\frac{J}2\int_0^1\left[\tanh\left(\tfrac{\be J\rho(x)}2\right)+
                \tanh\left(\tfrac{\be J\rho(x)}4\right)\right]\rho(x)\diff x,\\
  \fl
  c_V^{(4|4)}(T)&=\frac{\be^2J^2}8\int_0^1\left[2\sech^2\left(\tfrac{\be J\rho(x)}2\right)
                  +\sech^2\left(\tfrac{\be J\rho(x)}4\right)\right]\rho^2(x)\diff x,\\
  \fl
  s^{(4|4)}(T)&=\int_0^1\left[\vphantom{\frac{\be J\rho(x)}4}
                \log\left(8\cosh\left(\tfrac{\be J\rho(x)}2\right)
                \cosh^2\left(\tfrac{\be J\rho(x)}4\right)\right)\right.\\
  \fl
              &\hphantom{\int_0^1\bigg[\log\left(4\cosh\left(\tfrac{\be J\rho(x)}2\right)\right)}
                \left.-\frac{\be J\rho(x)}2
                              \bigg(\tanh\left(\tfrac{\be J\rho(x)}2\right)+
                   \tanh\left(\tfrac{\be J\rho(x)}4\right)\bigg)\right]\diff x.
\end{eqnarray*}
In particular, note that the previous formulas are obviously consistent with the general
relations~\eqref{fmm}--\eqref{smm}. See Fig.~\ref{fig.sumnexact} for a plot of the main
thermodynamic functions for the truly supersymmetric cases in which these functions have the
closed-form expressions listed above.

\begin{figure}[t]
  \centering
  \includegraphics[width=.48\linewidth]{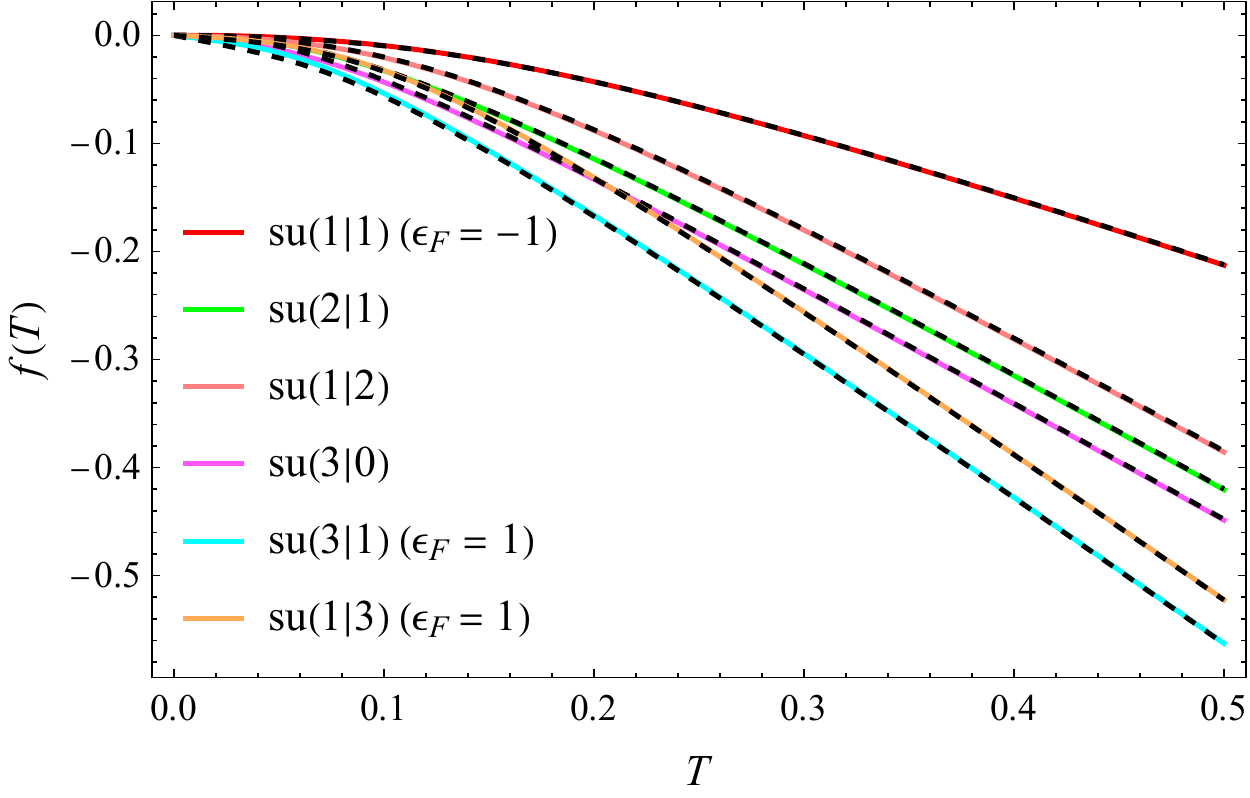}
  \caption{Free energy per spin of the $\su(1|1)$ (with $\vep=-1$), $\su(2|1)$, $\su(1|2)$,
    $\su(3|0)$, $\su(3|1)$ and $\su(1|3)$ chains~\eqref{Hnorm} (with $\vep_F=1$ in the last two
    cases) for $N=14$ spins (solid lines) vs.~the temperature $T$ (in units of $J$). The dashed
    lines represent the thermodynamic free energy per spin of the $\su(1|1)$, $\su(2|1)$,
    $\su(1|2)$, and $\su(3|0)$ HS chains, as well as the free energy per spin of the $\su(3|1)$
    and $\su(1|3)$ HS chains (whose thermodynamic limit is not explicitly known) for $N=14$
    spins.}
  \label{fig.conj}
\end{figure}

\begin{remark}
  We have noted above that the thermodynamic free energy per spin of the $\su(2|2)$, $\su(m|0)$,
  and $\su(0|m)$ (with even $m$) chains~\eqref{Hnorm} coincides with the free energy of their HS
  counterparts with the coupling $J$ rescaled to $2J$. Equivalently, by Remark~\ref{rem.1}, the
  free energy per spin of the latter chains does not change if we replace the spin flip operators
  $S_i$ by (plus or minus) the identity. It is also apparent from Fig.~\ref{fig.sumn4} that the
  thermodynamic free energies per spin of the $\su(3|1)$ and $\su(1|3)$ chains with $N=14$ spins
  are practically independent of $\vep_F$. These facts suggest that, in general, in the
  thermodynamic limit the free energy per spin of the $\su(m|n)$ chains~\eqref{Hnorm} with
  arbitrary $m$ and $n$ is independent of the representation of the spin reversal operators
  chosen. In particular, if this conjecture holds the free energy of the chain~\eqref{Hnorm}
  should coincide with the free energy of its HS counterpart with $J$ replaced by $2J$, obtained
  when $S_i$ is set to (plus or minus) the identity in Eq.~\eqref{Hnorm}. Although, as remarked
  above, the thermodynamic free energy per spin of the $\su(m|n)$ supersymmetric HS chain is not
  known exactly except for relatively low values of $m$ and $n$, we have numerically verified the
  latter conjecture in a few more cases by comparing the free energy per spin of both the
  chain~\eqref{Hnorm} and its HS counterpart (with $J$ replaced to $2J$) for a finite number of
  spins; see, e.g., Fig.~\ref{fig.conj}.\qed
\end{remark}

\begin{figure}
  \centering
    \includegraphics[width=.48\linewidth]{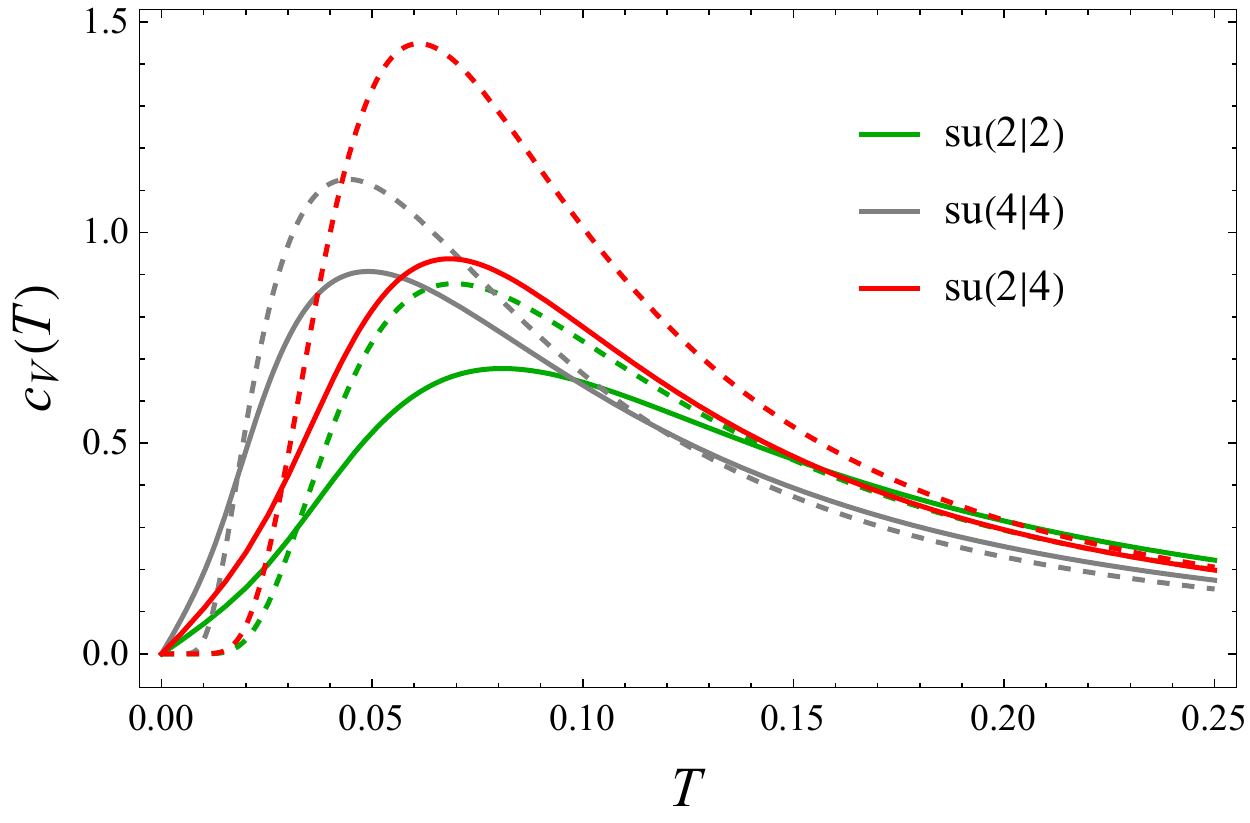}
    \caption{Specific heat per spin of the $\su(2|2)$, $\su(4|4)$, and $\su(2|4)$
      chains~\eqref{Hnorm} (solid curves), compared to their $k$-level counterparts (dashed
      curves). In all plots the temperature is measured in units of $J$.}
  \label{fig.cVmlevel}
\end{figure}

It is apparent from Figs.~\ref{fig.sumn4}-\ref{fig.sumnexact} that the qualitative behavior of the
thermodynamic functions of all the chains discussed in this work is very similar. In particular,
the specific heat per spin of all of these models features a single Schottky peak, typical of many
$k$-level systems. In the $\su(m|0)$ ---or, equivalently, $\su(n|0)$--- case this fact can be
explained by noting that the free energy~\eqref{fm0} of the $\su(m|0)$ chain can be written as
\[
  f^{(m|0)}(T)=-T\int_0^1\log\left(\sum_{k=0}^{m-1}\e^{-\frac{2kJ\be\rho(x)}{m}}\right)\diff x\,.
\]
As noted in Ref~\cite{FG22pre}, replacing $\rho(x)$ by its mean value over the interval $[0,1]$,
$1/6$, this partition function reduces to that of an $m$-level system with equally spaced,
non-degenerate levels with energies
\begin{equation*}
  E_i=\frac{iJ}{3m},\qquad i=0,\dots,m-1.
\end{equation*}
This suggests that the specific heat of the latter $m$-level system should qualitatively behave as
the specific heat per spin of the $\su(m|0)$ chain~\eqref{Hnorm}. That this is indeed the case can
be seen from Fig.~\ref{fig.cVmlevel}. This result also follows from the discussion of
Ref.~\cite{FG22pre}, since as noted above the thermodynamic functions of the $\su(m|0)$
chain~\eqref{Hnorm} coincide with those of the ferromagnetic $\su(m)$ $HS$ chain up to a trivial
rescaling of the coupling $J$. More interestingly, a similar argument can be applied to the
$\su(2|2)$ and $\su(4|4)$ chains. Indeed, in the former case the thermodynamic free energy per
spin can be expressed as
\[
  f^{(2|2)}(T)=-T\int_0^1\log\Bigl(1+2\e^{-\be J\rho(x)}+\e^{-2\be J\rho(x)}\Bigr)\diff x,
\]
which is expected to behave qualitatively as the free energy of a $3$-level system with energies
$E_k=kJ/6$ (with $k=0,1,2$) and degeneracies $g_0=g_2=1$, $g_1=2$. The specific heat of the this
three-level system, namely
\[
  c_{V}=\frac{\be^2J^2}{72}\sech^2\left(\be J/12\right),
\]
is again in reasonable agreement with that of the $\su(2|2)$ chain~\eqref{Hnorm}
(cf.~Fig.~\ref{fig.cVmlevel}). In particular, the temperature of the Schottky peak of the three-level
system, $T_{m}\simeq0.0695\,J$, is of the same order of magnitude as the corresponding
temperature for the $\su^{(2|2)}$ chain~\eqref{Hnorm}, $T_m^{(2|2)}\simeq0.0809\,J$. Similarly,
expressing the free energy per spin of the $\su(4|4)$ chain as
\begin{eqnarray*}
  \fl
  f^{(4|4)}(T)&=-T\int_0^1\log\left[(1+\e^{-\be J\rho(x)})(1+\e^{-\be J\rho(x)/2})^2)\right]\diff
                x\\
  \fl
  &=-T\int_0^1\log\Bigl(1+2\e^{-\be J\rho(x)/2}+2\e^{-\be J\rho(x)}+2\e^{-3\be J\rho(x)/2}
    +\e^{-2\be J\rho(x)}\Bigr)\diff x
\end{eqnarray*}
and replacing again $\rho(x)$ by its average over the interval $[0,1]$ we obtain a $5$-level
system with energies $E_k=kJ/12$ (with $k=0,\dots,4$) and degeneracies $g_0=g_4=1$,
$g_1=g_2=g_3=2$, whose specific heat is given by
\[
  c_{V}=\frac{\be^2}{72}\,\e^{-\be J/12}\,\frac{1+2\e^{-\be J/12}+2\e^{-\be J/6}+2\e^{-\be
      J/4}+\e^{-\be J/3}}{(1+\e^{-\be J/12})^2(1+\e^{-\be J/6})^2}.
\]
This specific heat provides again a reasonable approximation to the specific heat per spin
$c_V^{(4|4)}$, particularly at high temperatures (cf.~Fig.~\ref{fig.cVmlevel}). Again, the
temperature of the Schottky peak of the $5$-level system, given by $T_{m}\simeq0.0443 J$, is in
excellent agreement with the analogous temperature $T_m^{(4|4)}\simeq0.0491 J$ for the
$\su{(4|4)}$ chain.

The situation is somewhat murkier for the $\su(2|4)$ and $\su(4|2)$ chains, whose thermodynamic
free energy does has a more complicated structure than its $\su(m|0)$, $\su(1|1)$, $\su(2|2)$ and
$\su(4|4)$ counterparts. In the former case, at low temperatures we can approximate the
thermodynamic free energy as
\begin{eqnarray*}
  f^{(2|4)}(T)&\sim-T\int_0^1\log\Bigl[\big(1+\e^{-\be J\rho(x)}\big)\big(1+3\e^{-\be
                J\rho(x)}\big) \Bigr]\diff x \\
  &=-T\int_0^1\log\Bigl(1+4\e^{-\be J\rho(x)}+3\e^{-2\be
    J\rho(x)}\Bigr]\diff x,
\end{eqnarray*}
whose associated $3$-level system has energies $iJ/6$ (with $i=0,1,2$) and degeneracies $g_0=1$,
$g_1=4$, $g_2=3$. The specific heat of this $3$-level system,
\[
  c_{V}=\frac{\be^2}{9}\,\e^{-2\be J/3}\,\frac{1+3\e^{-\be J/6}+3\e^{-\be J/3}}{(1+4\e^{-\be
      J/6}+3\e^{-\be J/3})^2},
\]
should therefore provide a rough approximation to the specific heat per spin $c_V^{(2|4)}$.
Although in this case the agreement between both specific heats is noticeably poorer than in the
previous cases, the temperature of the Schottky peak of the three-level model,
$T_{m}\simeq0.0611 J$, is remarkably close to its counterpart $T_m^{(2|4)}\simeq0.0684 J$.

\section{Critical behavior}\label{sec.CB}

In this section we shall briefly analyze the critical behavior of the supersymmetric
chains~\eqref{Hnorm}. As is well known, one of the characteristic features of a conformal field
theory (CFT) is that its ground state (i.e., the vacuum) is non-degenerate, and that its spectrum
contains a subset of low-lying gapless excitations above the ground state with a linear
energy-momentum dispersion relation.
%
%
These two fundamental properties are thus a necessary condition for a finite-dimensional quantum
system to be critical, i.e., for its low-energy sector to be described by a CFT in the
thermodynamic limit. More precisely, the ground state of such a system must have at most a finite
degeneracy\footnote{By ``finite'' we mean independent of the number of particles in the
  thermodynamic limit. If the ground state of a critical system is $g$ times degenerate, its
  low-energy excitations will be described by $g$ identical copies of a single CFT.},
%
%
and its low-energy excitations must be gapless and feature a linear energy-momentum relation. We
shall therefore start by identifying in what cases the chain~\eqref{Hnorm} satisfies these
necessary conditions for criticality.
\begin{remark}
  Since (as noted in the previous section) we can assume without loss of generality that the
  coupling $J$ is positive, and the criticality properties of the chain~\eqref{Hnorm} are
  obviously independent of the size of $J$, for the sake of conciseness we shall set $J=1$
  throughout this section.\qed
\end{remark}

To begin with, we shall analyze for what values of $m$ and $n$ the ground state of the $\su(m|n)$
chain~\eqref{Hnorm} has a finite degeneracy. For $m\ne0$, this problem can be solved using the
explicit formulas for the ground state degeneracy $g$ of the ground state. From these formulas it
is straightforward to show that, as is the case with the ordinary supersymmetric HS chain, this
degeneracy cannot remain finite as $N\to\infty$ unless $m=1,2$. Indeed, for $m>2$ and $n$ both
even by Eq.~\eqref{dGSee} we have
\[
  g>2\sum_{k=0}^N(k+1)\binom{\frac{n}2}{N-k}\ge2(N+1).
\]
Similarly, for $m>2$ even and $n$ odd Eq.~\eqref{GSdegEO} implies that
\[
  g>\sum_{k=0}^N(k+1)\left[\binom{\frac12(n-1)}{N-k}+\binom{\frac12(n+1)}{N-k}\right]\ge2(N+1).
\]
The case $m>2$ odd and $n$ even is analogous, namely (by Eq.~\eqref{dGSoe})
\[
  g>\sum_{k=0}^N(k+2)\binom{\frac{n}2}{N-k}\ge N+2.
\]
Finally, when $m>2$ and $n$ are both odd Eq.~\eqref{GSdegOO} yields
\[
  g>\sum_{k=0}^N
  \left[(k+1)\binom{\frac12(n+\vep_F)}{N-k}+\binom{\frac12(n-\vep_F)}{N-k}\right]\ge N+2.
\]

Let us next study the ground state degeneracy for the cases $m=1,2$. For $m=2$, this degeneracy is
given by
\begin{equation}\label{gmtwo}
  g=\cases{2\sum_{k=0}^N\binom{\frac{n}2}{k}\le2^{\frac{n}2+1},& $n$ even\\
    \sum_{k=0}^N\left[\binom{\frac12(n-1)}{k} +\binom{\frac12(n+1)}{k}\right]\le
    3\cdot2^{\frac12(n-1)},& $n$ odd,}
\end{equation}
which is indeed finite (note that the equality holds for $N\ge n/2$ if $n$ is even or
$N\ge(n-1)/2$ if $n$ is odd). The case $m=1$ is dealt with analogously. Indeed,
\[
  \fl
  g=\cases{\binom{\frac{n}2}N+\sum_{k=0}^N\binom{\frac{n}2}{k}\le
    \binom{\frac{n}2}N+2^{\frac{n}2},& $n$ even\\
    \binom{\frac12(n-\vep_F)}N+\sum_{k=0}^N\binom{\frac12(n+\vep_F)}{k}\le
    \binom{\frac12(n-\vep_F)}N+2^{\frac12(n+\vep_F)},& $n$ odd,}
\]
which is finite (note that in this case the ground state degeneracy is simply $2^{\frac n2}$ for
$n$ even and $2^{\frac12(n+\vep_F)}$ for $n$ odd, if $N$ is respectively greater than $n/2$ or
$(n-\vep_F)/2$).

From the previous analysis it follows that the chain~\eqref{Hnorm} can only be critical for
$m=0,1,2$. The determination of the ground state degeneracy of the $\su(0|n)$ chain, or of the
existence of low-lying levels with a linear energy-momentum relation in all of the latter cases,
can be carried out with the results of the previous sections only when $m$ and $n$ are both even,
since only in this case the partition function can be expressed in terms of the partition
functions of the $\su(1|1)$ and $\su\bigl(\frac m2|\frac n2\bigr)$ HS chains. In particular, the
criticality of the $m=1$ case ---more precisely, the existence of low-lying levels with a linear
energy-momentum dispersion relation--- cannot be ascertained with the techniques of this paper,
and thus remains an open problem.

When $m$ and $n$ are both even, the equivalence of the chain Hamiltonian~\eqref{Hnorm} to the sum
of two non-interacting $\su(1|1)$ and $\su\bigl(\frac{m}2|\frac{n}2\bigr)$ HS chains
(cf.~Eq.\eqref{HHS}) entails that the energy spectrum of our model can be expressed in terms of
$\su(1|1)$ and $\su\bigl(\frac{m}2|\frac{n}2\bigr)$ bond vectors and their corresponding
(supersymmetric) motifs~\cite{BBH10}. More precisely, the spectrum of the chain~\eqref{Hnorm} with
$m$ and $n$ even can be generated from the formula
\begin{equation}\label{specmot}
  E(\bsi,\bsi')=\frac{1}{N^2}
  \sum_{i=1}^{N-1}\left[\de(\si_i,\si_{i+1})+\de'(\si'_i,\si'_{i+1})\right]i(N-i),
\end{equation}
where $\bsi\in\{1,2\}^N$, $\bsi'\in\{1,,\dots,(m+n)/2\}^N$ are respectively $\su(1|1)$ and
$\su\bigl(\frac{m}2|\frac{n}2\bigr)$ bond vectors, and the functions $\de$ and $\de'$ are defined
by
\[
  \de(j,k)=\cases{0,& $j<k$\en or\en $j=k=1$\\
  1,& $j>k$\en or\en $j=k=2$}
\]
and
\[
  \de'(j,k)=\cases{0,& $j<k$\en or\en $j=k\in\{1,\dots,\tfrac m2\}$\\
    1\vrule width0pt height16pt,& $j>k$\en or\en $j=k\in\{\tfrac m2+1,\dots,\tfrac{m+n}2\}$.}
\]
Equation~\eqref{specmot} implies that the ground state of the supersymmetric chain~\eqref{Hnorm}
with even $m$ and $n$ is obtained from the bond vectors $\bsi$ and $\bsi'$ yielding respectively
the ground states of the $\su(1|1)$ and $\su\bigl(\frac{m}2|\frac{n}2\bigr)$ HS chains. In
particular, the degeneracy of this ground state is the product of the degeneracies of the ground
states of the $\su(1|1)$ and $\su\bigl(\frac{m}2|\frac{n}2\bigr)$ chains. Since the ground state
of the $\su(1|1)$ HS chain is obviously obtained from the two bond vectors $(1,\dots,1,s)$ with
$s=1,2$, the previous observation entails that the ground state degeneracy of the $\su(m|n)$
chain~\eqref{Hnorm} with even $m$ and $n$ is twice the degeneracy of the ground state of the
$\sumnhalf$ HS chain. For $m=2$ and $N\ge n/2$, the ground state of the latter chain is the zero
mode obtained from an $\su(1|\frac{n}2)$ bond vector of the form
\[
  (\,\underbrace{1,\dots,1}_{N-k}\,,s_1,\dots,s_k),
\]
with $2\le s_1<\cdots<s_k\le n/2$ and $k\le n/2$. Since the number of such bond vectors is
$2^{n/2}$, the ground state degeneracy of the $\su(2|n)$ chain~\eqref{Hnorm} with even $n$ and
$N\ge n/2$ is $2^{\frac{n}2+1}$, in agreement with the remark after Eq.~\eqref{gmtwo}. Moreover,
it is known~\cite{BBS08,FGLR18} that both the $\su(1|1)$ and the $\su\bigl(1|\frac{n}2\bigr)$
chains have low energy excitations obtained by slightly varying the $\su(1|1)$ and
$\su\bigl(1|\frac{n}2\bigr)$ bond vectors near their ends. The typical energy of one of these
excitations is thus
\[
  \De E=\frac{i}{N^2}(N-i),
\]
with $i\ll N$, and its momentum is also known to be $\De p=\pm2\pi i/N$~\cite{BBS08}. Thus the
Fermi velocity of the low energy excitations of both the $\su(1|1)$ and
$\su\bigl(1|\tfrac{n}2\bigr)$ HS chains is
\[
  v_F=\lim_{N\to\infty}\left|\frac{\De E}{\De p}\right|=\frac1{2\pi}.
\]
Combining these excitations we obviously obtain low energy excitations above the ground state of
the $\su(2|n)$ chain~\eqref{Hnorm} with a finite Fermi velocity $1/(2\pi)$, so that this chain
does fulfill the two necessary conditions for criticality mentioned at the beginning of this
section. In fact, it is shown in Ref.~\cite{BBS08} that in the thermodynamic limit the low energy
excitations of the $\su(1|p)$ HS chains are described by a CFT of $p$ non-interacting massless
Majorana fermion fields. It follows that the $\su(2|n)$ chain~\eqref{Hnorm} is indeed critical,
with central charge
\begin{equation}\label{c2n}
  c^{(2|n)}=\frac12+\frac{n}4=\frac14(n+2).
\end{equation}

A similar analysis can be performed for the $\su(0|n)$ chain~\eqref{Hnorm} with even $n$. The main
difference is that in this case the $\su\bigl(0|\frac{n}2\bigr)$ HS chain contains no bosons, and
thus (for $N>n/2$) its ground state has positive energy. As explained in Ref.~\cite{FG22pre}, the
ground state in this case is obtained from $\su\bigl(0|\frac{n}2\bigr)$ bond vectors of the form
\[
  \big(1,2,\dots,\tfrac{n}2,\dots,1,2,\dots,\tfrac{n}2,s_1,\dots,s_{N-rn/2}\big)
\]
or their reflected analogues
\[
  \big(s_1,\dots,s_{N-rn/2},1,2,\dots,\tfrac{n}2,\dots,1,2,\dots,\tfrac{n}2\big),
\]
where $r=\lfloor 2N/n\rfloor$ and
\[
  1\le s_1<\cdots<s_{N-rn/2}\le \frac{n}2.
\]
Hence the ground state degeneracy of the $\su\bigl(0|\frac{n}2\bigr)$ HS chain is equal to
\[
  \cases{1,& $2N=nr$\\
    2\binom{\frac{n}2}{N-\frac{rn}2},& $2N>nr$,}
\]
and the degeneracy of the ground state of the $\su\bigl(0|\frac{n}2\bigr)$ chain~\eqref{Hnorm} is
twice the above number. In particular, the ground state degeneracy of the latter model remains
finite as $N\to\infty$. Furthermore, the $\su\bigl(0|\frac{n}2\bigr)$ HS chain is also known to
possess low energy excitations with finite Fermi velocity $v_F=1/(2\pi)$, described by the
$\su(n/2)_1$ WZNW model with central charge $(n/2)-1$~\cite{BBS08}. Reasoning as above we deduce
that the $\su(0|n)$ chain~\eqref{Hnorm} is critical, with central charge
\begin{equation}\label{c0n}
  c^{(0|n)}=\frac12+\frac{n}2-1=\frac12(n-1).
\end{equation}

Summarizing, we have just shown that the $\su(0|n)$ and $\su(2|n)$ chains~\eqref{Hnorm} with even
$n$ are critical, while the $\su(m|n)$ chains with $m>2$ are not. Moreover, the $\su(1|n)$ chains
with arbitrary $n$ and the $\su(2|n)$ chains with odd $n$ have been shown to have a finite ground
state degeneracy, and could thus be critical. On the other hand, with the results of this paper
nothing can be concluded about the $\su(0|n)$ chains with odd $n$, although our numerical
calculations suggest that these chains have a non-degenerate ground state and therefore could be
critical. We briefly sum up the above results in Table~\ref{tab.table}.
\begin{table}[h!]
  \centering
  \begin{tabular}[center]{|c|c|c|c|c|}
    \hline
    $(m,n)$& Finite GS degeneracy& $c$ & Associated CFT\\
    \hline\noalign{\vskip1.5pt}\hline
    $(0,2p)$& $\checkmark$& $p-\frac12$& 1 MMF~+ $\su(p)_1$ WNZW\\
    \hline
    $(0,2p-1)$& $*$ & ? & ?\\
    \hline
    $(1,n)$& $\checkmark$ & ? & ?\\
    \hline
    $(2,2p)$ &$\checkmark$ & $\frac12(p+1)$ & $p+1$ MMFs\\
    \hline
    $(2,2p-1)$ & $\checkmark$ & ? & ?\\
    \hline
  \end{tabular}
  \caption{Summary of results on the critical character of the $\su(m|n)$ supersymmetric
    chain~\eqref{Hchain}. Here $p$ and $n$ are positive integers, and the abbreviations GS and MMF
    stand respectively for ``ground state'' and ``massless Majorana fermion''. The asterisk ($*$)
    indicates that the finite degeneracy of the ground state of the $\su(0|2p-1)$ chain has been
    verified only for particular values of $p$ and the number of spins $N$, while the
    interrogation mark (?) stands for ``not known''. As shown above, the chains with $m\ge3$ have
    infinite ground state degeneracy in the thermodynamic limit, and cannot thus be critical.}
\label{tab.table}
\end{table}

As remarked in the Introduction, at low temperature the free energy of a critical quantum system
should have the same asymptotic behavior as the free energy of the CFT describing its low energy
sector, given by Eq.~\eqref{fcrit}. In particular, from the growth of the free energy of a quantum
critical system at low temperatures one can infer the central charge of its associated CFT, and
thus identify the system's universality class. Since the free energy of the $\su(m|n)$
chains~\eqref{Hnorm} with $m=0,2$ and $n$ even discussed above obeys Eq.~\eqref{fmnfHS}, it should
verify Eq.~\eqref{fcrit} with $v_F=1/(2\pi)$ (the Fermi velocity of the critical supersymmetric HS
chains) and central charge
\begin{equation}\label{centc}
  c^{(m|n)}=c^{(1|1)}_{\mathrm{HS}}+c^{(\frac{m}2|\frac
    n2)}_{\mathrm{HS}}=\frac12+c^{(\frac{m}2|\frac n2)}_{\mathrm{HS}}.
\end{equation}
This formula is in fact in full agreement with the result obtained above studying the behavior of
the low-lying energy excitations (cf.~Eqs.~\eqref{c2n} and \eqref{c0n}).

The low temperature behavior~\eqref{fcrit} of the free energy per spin of the $\su(0|n)$ chain
with even $n$, the $\su(2|2)$ and the $\su(2|4)$ chains (and, in particular, Eq.~\eqref{centc} for
the central charge) can be explicitly checked using the formulas for their thermodynamic free
energy per spin derived in the previous section. Indeed, in all of these cases the free energy can
be expressed in terms of integrals of the form
\begin{equation}\label{Ibe}
  I_\be[\vp]:=\int_0^{1/2}\diff x \,\log\Bigl(\vp(\e^{-\be\rho(x)})\Bigr),
\end{equation}
where $\vp(z)$ is a smooth function such that $\vp(z)>0$ for $z>0$, $\vp(0)=1$ and\footnote{Note
  that that $z^\al=O(z)$ as $z\to0+$ for $\al\ge1$.}
\begin{equation}\label{vpasy}
  \vp(z))=1+O(z)
\end{equation}
as $z\to0+$.
Using Eq.~\eqref{Ibevp} in~\ref{app.asint} to approximate this integral, it is straightforward to
derive the asymptotic behavior at low temperatures of the thermodynamic free energy per spin of
the $\su(0|n)$ (with even $n$), $\su(2|2)$ and $\su(2|4)$ chains. To begin with, using
Eqs.~\eqref{dualrel} and~\eqref{fm0} (with $J=1$) the free energy of the $\su(0|n)$
chain~\eqref{Hnorm} with $n$ even can be expressed as
\begin{eqnarray*}
  f^{(0|n)}(T)&=\frac{1}6\left(1+\frac1{n}\right)-T\int_0^1\diff x\,\log \left(\frac{\sinh\Big(\be
                \rho(x)\Big)}{\sinh\Big(\be \rho(x)/n\Big)}\right)\\
              &=\frac{1}{3n}
                -T\int_0^1\diff x\,\log\left(\frac{1-\e^{-2\be \rho(x)}}{1-\e^{-2\be \rho(x)/n}}\right).
\end{eqnarray*}
Taking into account that $\rho(x)$ is symmetric about $x=1/2$ we can rewrite the previous formula
as
\[
  f^{(0|n)}(T)=f^{(0|n)}(0)-2T\left(I_\be[1-z^2]-I_{\be/n}[1-z^2]\right).
\]
From Eq.~\eqref{Ibevp} we then have
\begin{eqnarray*}
  \fl
  I_\be[1-z^2]&=T\int_0^\infty\diff y\,\log(1-\e^{-2y})+O(T^2)=\frac{T}2\int_0^\infty\diff y\,\log(1-\e^{-y})+O(T^2)
  \\ \fl
              &=-\frac{T}2\ze(2)+O(T^2)=-\frac{\pi^2T}{12}+O(T^2),
\end{eqnarray*}
and therefore
\[
  f^{(0|n)}(T)-f^{(0|n)}(0)=-(n-1)\frac{\pi^2T^2}{6}+O(T^3)=-\frac12(n-1)\frac{\pi T^2}{6v_F}
  +O(T^3).
\]
Comparing with Eq.~\eqref{fcrit} we conclude that in this case the central charge is given by
Eq.~\eqref{c0n}. Likewise, by Eq.~\eqref{su22} with $J=1$ the thermodynamic free energy of the
$\su(2|2)$ chain is given by
\[
  f^{(2|2)}(T)=2f^{(1|1)}_{\mathrm{HS}}(T)=-4T I_\be[1+z],
\]
and thus
\begin{eqnarray*}
  \fl
  f^{(2|2)}(T)&=-4T^2\int_0^\infty\diff y\,\log(1+\e^{-y})+O(T^3)=-2T^2\ze(2)+O(T^3)
                \\
  \fl
  &=-\frac{\pi^2T^2}3+O(T^3)=-\frac{\pi T^2}{6v_F}+O(T^3).
\end{eqnarray*}
Hence the central charge of the $\su(2|2)$ chain is $c^{(2|2)}=1$, in agreement with
Eq.~\eqref{c2n}. Finally, by Eq.~\eqref{fsu24} with $J=1$ we have
\[
  f^{(2|4)}(T)=-2T I_\be[1+z]-2T I_\be\Bigl[\tfrac12+z+\tfrac12\sqrt{1+8z}\,\Bigr],
\]
and therefore
\begin{eqnarray*}
  \fl
  f^{(2|4)}(T)&=-\frac{\pi^2T^2}{12}-2T^2\int_0^\infty\diff y\,\log
  \Bigl(\tfrac12+\e^{-y}+\tfrac12\sqrt{1+8\e^{-y}}\,\Bigr)
                +O(T^3)\\
  \fl
              &=-\frac{\pi^2T^2}{6}-\frac{\pi^2T^2}{3}+O(T^3)=-\frac{\pi^2T^2}2+O(T^3)=
                -\frac32\,\frac{\pi T^2}{6v_F}+O(T^3),
\end{eqnarray*}
so that the central charge of the $\su(2|4)$ chain is again given by Eq.~\eqref{c2n} with $n=4$.
\begin{figure}[t]
  \centering
  \includegraphics[width=.48\linewidth]{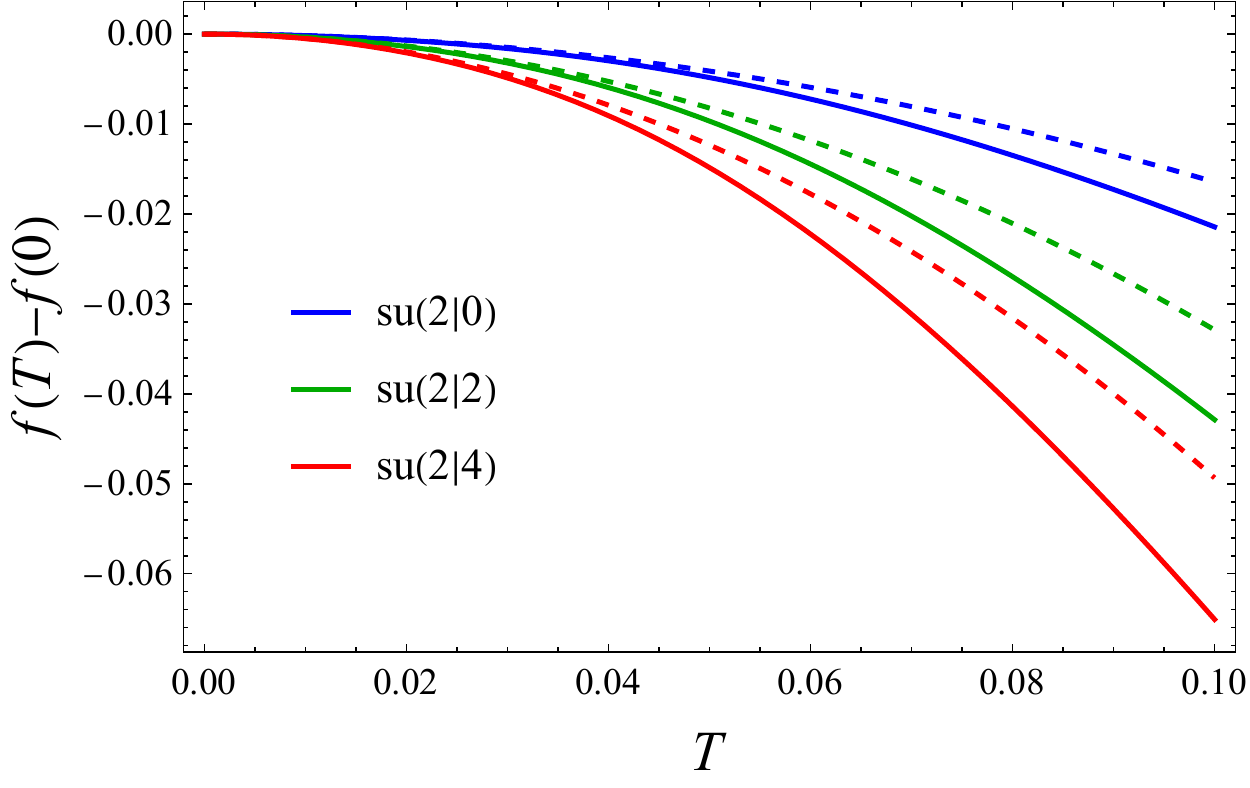}
  \caption{Thermodynamic free energy per spin of the $\su(0|2)$, $\su(2|2)$ and $\su(2|4)$
    critical chains (solid line) compared to their low temperature approximations (dashed curves).
    As usual, the temperature is measured in units of $J$.}
  \label{fig.ffapp}
\end{figure}
This concludes the verification of Eq.~\eqref{centc} for the $\su(0|n)$, $\su(2|2)$ and $\su(2|4)$
chains (see Fig.~\ref{fig.ffapp} for a plot comparing $f(T)-f(0)$ for the latter chains with its
low temperature approximations derived above).

\section{Conclusions and outlook}\label{sec.conc}

In this paper we have introduced a novel family of $\su(m|n)$ supersymmetric, translationally
invariant spin chains with long-range interactions. The new chain's spin-spin interaction term,
which depends on supersymmetric permutation and spin reversal operators, reduces to the
interaction term of the celebrated (supersymmetric) Haldane--Shastry chain when the spin reversal
operators are replaced by (plus or minus) the identity. On the other hand, by contrast with all
previously known spin chains of Haldane--Shastry type, the new model is not directly associated
with an extended root system. We show that the spin chain under study can be obtained from a
suitable many-body spin dynamical model in the strong coupling limit, and take advantage of this
fact to evaluate the chain's partition function in closed form. The structure of the partition
function turns out to depend crucially on the parity of the integers $m$ and $n$. In particular,
we show that when both $m$ and $n$ are even the partition function exactly factorizes as the
product of the partition functions of an $\su(1|1)$ and an $\sumnhalf$ Haldane--Shastry spin
chains. We also study in detail the chain's symmetries, showing that it features a remarkable
invariance under translations combined spin ``twists'' (i.e., spin flips at one end of the chain),
as well as a general boson-fermion duality as is the case with supersymmetric spin chains of
Haldane--Shastry type. In fact, the structure of the partition mentioned above when both $m$ and
$n$ are even strongly suggests that in this case the model admits the direct sum of the Yangians
$Y(\gl(1|1))$ and $Y(\gl\bigl(\tfrac{m}2|\tfrac{n}2\bigr))$ as an exact symmetry for an arbitrary
number of sites.

The explicit knowledge of the new chain's partition function makes it possible to study its
thermodynamics in a systematic way. More precisely, when both $m$ and $n$ are even we are able to
find a closed-form expression for the thermodynamic free energy per spin for low values of $m$ and
$n$, and for arbitrary even $m$ or $n$ in the non-supersymmetric case $mn=0$. For other values of
$m$ and $n$, we compute the free energy per spin for a finite number of spins from the partition
function and show that the thermodynamic functions behave similarly as in the above mentioned
cases. In particular, the specific heat at constant volume exhibits a single marked Schottky peak,
whose appearance can be understood by noting the qualitative similarity of the chain's free energy
with the free energy of a suitable $k$-level system. Our results also have led us to conjecture
that in the thermodynamic limit the free energy per spin is independent of the particular
representation of the spin reversal operators $S_i$ used. In particular, this result applied to
the trivial representation $S_i=\pm\id$ would imply that the thermodynamic free energy per spin of
the new chain exactly coincides with its analogue for the corresponding HS chain with the coupling
$J$ doubled. This equality has been in fact verified, either exactly or numerically, in several
cases.

We have also studied in detail the critical behavior of the new chain introduced in this paper. To
begin with, from the ground state degeneracy we have been able to show that the model cannot be
critical when $m>2$. With the help of the closed-form expression for the partition function, we
have established the criticality of the chains with $m=0,2$ and even $n$. We have also checked
that at low temperatures the free energy per spin of the $\su(0|n)$ (with even $n$), $\su(2|2)$,
and $\su(2|4)$ chains behaves as expected for a critical system, and have computed the central
charge of their associated CFTs.

The above results suggest several open problems and related lines for future work. In the first
place, it would be desirable to find a simple description of the spectrum when either $m$ or $n$
(or both) are odd in terms of some variant of the supersymmetric motifs introduced in
Section~\ref{sec.CB}. A potential candidate for such a variant could be the branched motifs
introduced in Ref.~\cite{BS20} to generate the spectrum of the supersymmetric Polychronakos--Frahm
(PF) spin chain of $BC_N$ type. Note that a motif-based description of the spectrum could shed
light on the existence of low energy excitations with a linear energy-momentum relation, which is
a necessary condition for criticality, for the $\su(0|n)$ and $\su(2|n)$ chains with odd $n$ and
the $\su(1|n)$ chain. Another problem worth investigating is the existence of a factorization of
the partition function of the $\su(m|n)$ chains with odd $m$ or $n$ in terms of the partition
functions of HS spin chains of $A_{N-1}$ type, akin to Eq.~\eqref{fact} for the case of even $m$
and $n$. This would automatically yield a motif-based description of the spectrum, and imply the
existence of a Yangian symmetry for arbitrary values of $m$ and $n$. More generally, it would
certainly be desirable to find a closed-form expression for the thermodynamic free energy per spin
valid for all values of $m$ and $n$. In particular, the low-temperature behavior of this
expression would shed light on the critical character of the $\su(0|n)$ and $\su(2|n)$ chains with
odd $n$, as well as the $\su(1|n)$ chain, since it could be used to compute the central charge of
the hypothetical CFTs associated to these models. In any case, rigorously determining the
criticality of the latter chains, and identifying their associated CFTs, are natural open problems
suggested by the present work. Finally, it would also be worthwhile to probe our conjecture on the
independence of the free thermodynamic energy per spin on the particular representation of the
spin reversal operators chosen, not only for the chains introduced in this work but also for
similar models (for instance, HS chains of $BC_N$ type).

\ack This work was partially supported by grant~GRFN24/24 from Universidad Complutense de Madrid.
The authors would like to thank Professors Diptiman Sen and Germán Sierra for useful discussions.

\appendix

\section{The $\su(m|n)$ superalgebra}\label{app.sumn}

In this appendix we shall spell out the precise connection between the $\su(m|n)$ Lie superalgebra
and the supersymmetric chain~\eqref{Hchain}. To begin with, we note that the basis
states~\eqref{basis} can be obtained from a Fock vacuum~$\ket{\vac}$ as
\[
  \ket{s_1\cdots s_N}=\dc_{1,s_1}\cdots \dc_{N,s_N}\ket{\vac},
\]
where $\dc_{i\al}$ creates a particle of type $\al$ (i.e., a boson for $\al=1,\dots,m$ and a
fermion for $\al=m+1,\dots,m+n$) at site $i$. The operators $c_{i\al}$, $\dc_{i\al}$ satisfy the
usual canonical (anti)commutation relations
\[
  [c_{i\al},\dc_{j\be}]_\pm:=c_{i\al}\dc_{j\be}-(-1)^{p(\al)p(\be)}=\de_{ij}\de_{\al\be},
\]
where the parity $p(\al)$ is defined as $0$ for bosons and $1$ for fermions. The chain's Hilbert
space coincides with the subspace of the Fock space defined by the conditions
\begin{equation}\label{constr}
  \sum_{\al=1}^{m+n}\dc_{k\al}c_{k\al}=\id_k,\qquad k=1,\dots,N,
\end{equation}
enforcing restriction that there be exactly one particle per site. It is straightforward to check
that the $\su(m|n)$ supersymmetric permutation and spin flip operators $S_{ij}$ and $S_i$
appearing in the Hamiltonian~\eqref{Hchain} can be expressed in terms of the operators
\begin{equation}
  \label{Ealbe}
  E_k^{\al\be}:=\dc_{k\al}c_{k\be},\qquad k=1,\dots,N,\quad \al,\be=1,\dots,m+n,
\end{equation}
as
\begin{equation}
  \label{SijSiE}
  S_{ij}=\sum_{\al,\be=1}^{m+n}(-1)^{p(\be)}E_i^{\al\be}E_j^{\be\al},\qquad
  S_{i}=\sum_{\al=1}^{m+n}\si(\al)E_i^{\al'\al},
\end{equation}
where the $\si$ and prime ($\,'\,$) notation was defined in Eqs.~\eqref{sigma}-\eqref{prime}. It
is also easily verified that the operators~\eqref{Ealbe} at each site $k=1,\dots,N$ realize the
commutation relations of the $\gl(m|n)$ Lie superalgebra, namely\cite{Ri78,JG79}
\[
  \fl
  [E^{\al\be},E^{\ga\de}]_\pm:=E^{\al\be}E^{\ga\de}-(-1)^{p(\al,\be)p(\ga,\de)}E^{\ga\de}E^{\al\be}
  =\de_{\be\ga}E^{\al\de}-(-1)^{p(\al,\be)p(\ga,\de)}\de_{\al\de}E^{\ga\be},
\]
where $p(\mu,\nu)=p(\mu)+p(\nu)$ is the parity of the generator~$E^{\mu\nu}$. Note that the linear
combination $\sum_{\al=1}^{m+n}E^{\al\al}$ is a linear Casimir of the $\gl(n|m)$ superalgebra,
since
\[
  \fl
  \bigg[E^{\al\be},\sum_{\ga=1}^{m+n}E^{\ga\ga}\bigg]_{\pm}=
  \bigg[E^{\al\be},\sum_{\ga=1}^{m+n}E^{\ga\ga}\bigg]
  =\sum_{\ga=1}^{m+n}\left(\de_{\be\ga}E^{\al\ga}-\de_{\al\ga}E^{\ga\be}\right)=E^{\al\be}-E^{\al\be}=0.
\]
For the representations~\eqref{Ealbe} of $\gl(m|n)$ in terms of creation and annihilation
operators at site $k$ this Casimir takes the value $1$ on account of the
constraint~\eqref{constr}. Let us next define the supertrace of an operator $A_k$ acting on the
$k$-th site as
\[
  \str A_k=\sum_{\al=1}^{m+n}(-1)^{p(\al)}{}_k\bra\al A\ket\al_k,
\]
where $\ket\al_k=\dc_{k\al}\ket\vac$. Clearly
\[
  \fl
  \str E^{\al\be}_k=\sum_{\ga=1}^{m+n}(-1)^{p(\ga)}\bra\vac c_{k\ga}^\ga\dc_{k\al}
  c_{k\be}\dc_{k\ga}\ket\vac=\sum_{\ga=1}^{m+n}(-1)^{p(\ga)}\de_{\ga\al}\de_{\be\ga}
  =(-1)^{p(\al)}\de_{\al\be}.
\]
In order to obtain a representation of the Lie superalgebra $\sla(m|n)$, we must impose the
additional condition that the supertrace of its generators vanish. When $m\ne n$, this is can be
done in a symmetric way by defining the new generators
\begin{equation}\label{Jalbe}
  J^{\al\be}_k=E^{\al\be}_k-\frac{(-1)^{p(\al)}}{m-n}\de_{\al\be}\id_k,
\end{equation}
since $\str\id_k=m-n$. Note that the dimension of $\sla(m|n)$ is $(m+n)^2-1$, since we have the
constraint
\begin{equation}\label{Jalal}
  \sum_{\al=1}^{m+n}J_k^{\al\al}=\sum_{\al=1}^{m+n}E_k^{\al\al}-\id_k=0
\end{equation}
by Eq.~\eqref{constr}. When $m=n$ it is still possible to construct the $\sla(m|m)$ generators out
of the $\gl(m|n)$ generators~\eqref{Ealbe} in a less symmetric way by defining, for instance,
$J_k^{\al\be}=E_k^{\al\be}$ for $\al\ne\be$ and
\[
  J^{\al\al}_k=(-1)^{p(\al)}E_k^{\al\al}-(-1)^{p(\al+1)}E_k^{\al+1,\al+1},\qquad
  \al=1,\dots,N-1.
\]
In fact, while the Lie superalgebra $\sla(m|n)$ is simple when $m\ne n$~\cite{Ka77AM,Ri78},
$\sla(m|m)$ is not. Indeed, when $m=n$ the one-dimensional linear space spanned by the Casimir
$\sum_{\al=1}^{2m}E_k^{\al\al}$ ---which belongs to $\sla(m|n)$ for $m=n$--- is obviously an ideal
of $\sl(m|m)$. For the sake of simplicity (as is usually done in the literature) we shall
therefore restrict ourselves in what follows to the case $m\ne n$.

When $m\ne n$, an elementary calculation shows that the $\sla(m|n)$ generators~\eqref{Jalbe} at
site $k$ satisfy exactly the same commutation relation as the $\gl(m|n)$ generators~\eqref{Ealbe},
namely
\[
  [J_k^{\al\be},J_k^{\ga\de}]_\pm
  =\de_{\be\ga}J_k^{\al\de}-(-1)^{p(\al,\be)p(\ga,\de)}\de_{\al\de}J_k^{\ga\be},\qquad
  \al,\be=1,\dots,m+n.
\]
Formally these are also the $\su(m|n)$ commutation relations, since $\sla(m|n)$ is the
complexification of $\su(m|n)$.

It is straightforward to express the operators $S_{ij}$ and $S_i$ in terms of the $\su(m|n)$
---or, strictly speaking, $\sla(m|n)$--- generators $J_k^{\al\be}$ at sites $i$ and $j$. Indeed,
using Eq.~\eqref{Jalal} we obtain
\begin{eqnarray*}
  \fl
  \sum_{\al,\be}(-1)^{p(\be)}J_i^{\al\be}J_j^{\be\al}
  &\equiv\sum_{\al,\be}(-1)^{p(\al)}(J_i^{\al\be})^\dagger J_j^{\al\be}\\
  \fl
  &=\sum_\al\left(\dc_{i\al}c_{i\al}-\frac{(-1)^{p(\al)}}{m-n}\,\id\right)(-1)^{p(\al)}J_j^{\al\al}
    +\sum_{\al\ne\be}(-1)^{p(\be)}\dc_{i\al}c_{i\be}\dc_{j\be}c_{j\al}\\
  \fl
  &=\sum_\al(-1)^{p(\al)}\dc_{i\al}c_{i\al}J_j^{\al\al}
    +\sum_{\al\ne\be}\dc_{i\al}\dc_{j\be}c_{i\be}c_{j\al}\\
  \fl
  &=\sum_\al(-1)^{p(\al)}\dc_{i\al}c_{i\al}\left(\dc_{j\al}c_{j\al}-\frac{(-1)^{p(\al)}}{m-n}\,\id\right)
    +\sum_{\al\ne\be}\dc_{i\al}\dc_{j\be}c_{i\be}c_{j\al}\\
  \fl
  &=\sum_{\al,\be}\dc_{i\al}\dc_{j\be}c_{i\be}c_{j\al}
    -\frac{\id}{m-n}\sum_\al\dc_{i\al}c_{i\al}=S_{ij}-\frac{\id}{m-n},
\end{eqnarray*}
and therefore
\begin{equation}\label{SijJ}
  S_{ij}=\sum_{\al,\be}(-1)^{p(\be)}J_i^{\al\be}J_j^{\be\al}+\frac{\id}{m-n}.
\end{equation}
Similarly,
\begin{eqnarray}
  S_i&=\sum_\al\si(\al)E_i^{\al'\al}=\sum_\al\si(\al)\left(J_i^{\al'\al}
       -\frac{(-1)^{p(\al)}}{m-n}\de_{\al\al'}\id\right)\nonumber\\
     &=\sum_\al\si(\al)J_i^{\al'\al}
       +\frac{\pi(n)\vep_F
       -\pi(m)\vep_B}{m-n}\,\id,
  \label{SiJs}
\end{eqnarray}
where $\pi(\al)=(1-(-1)^\al)/2$ is the parity of the integer $\al$. Since $\tS_{ij}=S_{ij}S_iS_j$,
it follows that the operators $\tS_{ij}$ appearing in the chain Hamiltonian~\eqref{Hchain} can be
expressed as a fourth degree polynomial in the $\su(m|n)$ generators~\eqref{Jalbe} at sites $i$
and $j$. Obviously the same is then true for the full Hamiltonian~\eqref{Hchain}, and it is
precisely in this sense that the latter model can be regarded as an $\su(m|n)$ supersymmetric
chain.

Although not strictly necessary, we can replace the non-Hermitian generators $J^{\al\be}_k$ in
Eq.~\eqref{Jalbe} by the Hermitian linear combinations
\begin{eqnarray*}
  A_k^{\al\be}&=\frac12\left(J_k^{\al\be}+J_k^{\be\al}\right),\quad 1\le\al\le\be\le m+n;\\
  B_k^{\al\be}&=\frac\iu2\left(J_k^{\be\al}-J_k^{\al\be}\right),\quad 1\le\al<\be\le m+n,
\end{eqnarray*}
which obey the constraint
\[
  \sum_{\al=1}^{m+n}A_k^{\al\al}\equiv\sum_{\al=1}^{m+n}J_k^{\al\al}=0.
\]
After a straightforward calculation we arrive at the relations
\begin{eqnarray*}
  \fl
  S_{ij}&=\sum_{\al=1}^{m+n} (-1)^{p(\al)}A_i^{\al\al}A_j^{\al\al}+
          2\sum_{r<s}\left(A_i^{rs}A_j^{rs}+B_i^{rs}B_j^{rs}\right)
          -2\sum_{\rho<\si}\left(A_i^{\rho\si}A_j^{\rho\si}+B_i^{\rho\si}B_j^{\rho\si}\right)\\
  \fl
        &\qquad+2\iu\sum_{r,\rho}\left(A_i^{r\rho}B_j^{r\rho}
          -B_i^{r\rho}A_j^{r\rho}\right)+\frac\id{m-n},\\
  \fl
  S_i&=2\vep_B\sum_{r\le m/2}A_{i}^{rr'}+2\vep_F\sum_{\rho\le m+n/2}A_{i}^{\rho\rho'}
       +\pi(m)\vep_B\left(A_i^{\frac{m+1}2 \,\frac{m+1}2}-\frac\id{m-n}\right)\\
  \fl
        &\qquad
          +\pi(n)\vep_F\left(A_i^{m+\frac{n+1}2 \,m+\frac{n+1}2}+\frac\id{m-n}\right),
\end{eqnarray*}
where $r$ and $s$ (respectively $\rho$ and $\si$) run over the bosonic indices $1,\dots,m$
(respectively the fermionic indices $m+1,\dots,m+n$).

\section{Asymptotic approximation of the integral~\eqref{Ibe}}\label{app.asint}

The asymptotic behavior of the integral $I_\be[\vp]$ in Eq.~\eqref{Ibe} as $\be\to\infty$ can be
ascertained performing the change of variable $\be\rho(x)=y$, or equivalently
\[
  x=\rho^{-1}(Ty)=\frac12\left(1-\sqrt{1-4Ty}\,\right),\qquad 0\le y\le \be/4,
\]
whence
\begin{eqnarray*}
  \fl
  I_\be[\vp]&=T\int_0^{\be/4}\diff
              y\,\frac{\log\left(\vp(\e^{-y})\right)}{\sqrt{1-4Ty}}\\
  \fl
            &= TI[\vp]-T\int_{\be/4}^\infty\diff y\,\log\left(\vp(\e^{-y})\right)
              +T\int_0^{\be/4}\diff
              y\,\left(\frac1{\sqrt{1-4Ty}}-1\right)\log\left(\vp(\e^{-y})\right)
\end{eqnarray*}
with
\[
  I[\vp]:=\int_0^\infty\diff y\,\log\bigl(\vp(\e^{-y})\bigr).
\]
Note that the assumption~\eqref{vpasy} on $\vp(z)$ implies that
\begin{equation}\label{logvpasy}
  \log(\vp(z))=\log\left(1+O(z)\right)=O(z)
\end{equation}
as $z\to0+$, from which it follows that $I[\vp]$ is convergent and
\[
  \int_{\be/4}^\infty\diff y\,\log\left(\vp(\e^{-y})\right)=O\left(\e^{-\be/4}\right).
\]
To cope with the (integrable) singularity of the integrand at the upper integration limit in the
integral
\[
\int_0^{\be/4}\diff
              y\,\left(\frac1{\sqrt{1-4Ty}}-1\right)\log\left(\vp(\e^{-y})\right),
\]
we note that
\[
  \frac1{\sqrt{1-4z}}-1=h'(z),\qquad
  \text{with}\quad h(z)=\frac12\left(1-2z-\sqrt{1-4z}\right).
\]
Taking into account that $h(0)=0$ and integrating by parts we obtain
\begin{eqnarray*}
  \fl
  \int_0^{\be/4}\diff y\,\left(\frac1{\sqrt{1-4Ty}}-1\right)\log\left(\vp(\e^{-y})\right)
  &=\int_0^{\be/4}\diff y\,h'(Ty)\log\left(\vp(\e^{-y})\right)\\ \fl
  &=\frac{1}{4T}\,\log\left(\vp(\e^{-\be/4})\right)
    +\frac1T\int_0^{\be/4}\diff y\,h(Ty)\,\e^{-y}\frac{\vp'(\e^{-y})}{\vp(\e^{-y})}.
\end{eqnarray*}
The first term is clearly $O(\be\e^{-\be/4})$ as $T\to0+$ by Eq.~\eqref{logvpasy}. On the other
hand, from the elementary inequalities
\[
  0\le h(z)\le 4z^2
\]
it follows that
\[
  \frac1T\bigg|\int_0^{\be/4}\diff y\,h(Ty)\,\e^{-y}\frac{\vp'(\e^{-y})}{\vp(\e^{-y})}\bigg|
  \le 4T\int_0^{\infty}\diff y\,y^2\e^{-y}\left|\frac{\vp'(\e^{-y})}{\vp(\e^{-y})}\right|=O(T),
\]
since the integral is convergent. Indeed, by Eq.~\eqref{logvpasy} we have
\[
  \e^{-y}\left|\frac{\vp'(\e^{-y})}{\vp(\e^{-y})}\right|=O(\e^{-y})
\]
as $y\to\infty$. Putting everything together we conclude that
\begin{equation}\label{Ibevp}
  I_\be[\vp]=T I[\vp]+O(T^2)
\end{equation}
as $T\to0+$.

\section*{References}


\end{document}